\documentclass{jfm}
\usepackage{graphicx}
\usepackage{color}

\newcommand{\Du}{\mbox{D\hspace{-0.5pt}u}}
\newcommand{\Pe}{\mbox{P\hspace{-0.5pt}e}}

\newcommand{\bJ}{\mbox{\bf J}}

\newcommand{\eps}{\varepsilon_w}
\newcommand{\etw}{\varepsilon_d}
\newcommand{\epz}{\varepsilon_0}

\newcommand{\ep}{\epsilon}

\newcommand{\Real}{\mbox{Re}}  

\def\bea{\begin{eqnarray}}
\def\eea{\end{eqnarray}}

\newcommand{\oo}[1]{{\cal O}\left( #1 \right)}

\newcommand{\dendif}{c^\rho}
\newcommand{\densum}{c^e}
\newcommand{\ddendif}{\dot{c}^\rho}
\newcommand{\ddensum}{\dot{c}^e}

\newcommand{\nnabla}{\tilde{\nabla}}
\newcommand{\nzeta}{\tilde{\zeta}}
\newcommand{\npsi}{\tilde{\psi}}
\newcommand{\nphi}{\tilde{\phi}}
\newcommand{\bnu}{\tilde{{\bf u}}}
\newcommand{\nn}{\tilde{n}}
\newcommand{\nl}{\tilde{l}}
\newcommand{\nt}{\tilde{t}}
\newcommand{\nr}{\tilde{{\bf r}}}
\newcommand{\ndendif}{\tilde{c}^\rho}
\newcommand{\ndensum}{\tilde{c}^e}
\newcommand{\nE}{\tilde{E}}

\newcommand{\hnabla}{\bar{\nabla}}
\newcommand{\hphi}{\bar{\phi}}
\newcommand{\hn}{\bar{n}}

\newcommand{\htt}{\bar{t}}
\newcommand{\hdendif}{\bar{c}^\rho}
\newcommand{\hr}{\bar{{\bf r}}}

\newcommand{\be}{\begin{equation}}
\newcommand{\ee}{\end{equation}}
\newcommand{\br}{{\bf r}}

\newcommand{\thh}{{\bf {\hat \theta}}}
\newcommand{\nh}{{\bf {\hat n}}}
\newcommand{\brh}{{\bf {\hat r}}}
\newcommand{\zh}{{\bf {\hat z}}}

\newcommand{\bu}{{\bf u}}

\newcommand{\bF}{{\bf F}}
\newcommand{\bU}{{\bf U}}

\def\ie{{\it i.e.\ }}
\def\eg{{\it e.g. }}

\newcommand{\bj}{{\bf j}}
\newcommand{\bnh}{{\bf \hat{n}}}

\newcommand{\bv}{{\bf v}}

\newcommand{\bE}{{\bf E}}

\newcommand{\z}{\zeta}

\newcommand{\pd}[2]{\frac{\partial #1}{\partial #2}}

\title[Induced-charge electro-osmosis]{Induced-charge electro-osmosis}

\author[T. M. Squires and M. Z. Bazant]{T\ls O\ls D\ls D\ns 
M.\ns S\ls Q\ls U\ls I\ls R\ls E\ls S$^1$\ns
  \and M\ls A\ls R\ls T\ls I\ls N\ns Z.\ns B\ls A\ls Z\ls A\ls N\ls T$^2$}

\affiliation{$^1$Departments of Applied and Computational Mathematics
and Physics,\\ California Institute of Technology, Pasadena, CA 91125 \\
$^2$Department of Mathematics and Institute for Soldier
Nanotechnologies,\\ Massachusetts Institute of Technology, Cambridge, MA
02139}

\pubyear{2003} \volume{?} \pagerange{?--?}
\date{\today}
\begin{document}

\maketitle

\begin{abstract}
We describe the general phenomenon of `induced-charge electro-osmosis' (ICEO)
 --- the nonlinear electro-osmotic slip that occurs when an applied
field acts on the ionic charge it {\sl induces} around a polarizable
surface. Motivated by a simple physical picture, we calculate ICEO flows around
conducting cylinders in steady (DC), oscillatory (AC), and suddenly-applied 
electric fields.  This picture, and these systems, represent perhaps the clearest 
example of nonlinear electrokinetic phenomena.  We complement and verify this 
physically-motivated approach using a matched asymptotic expansion to the 
electrokinetic equations in the thin double-layer and low potential limits.
ICEO slip velocities vary like $u_s \propto E_0^2 L$, where $E_0$ is the field 
strength and $L$ is a geometric length scale, and are set up on a time scale $\tau_c =
\lambda_D L/D$, where $\lambda_D$ is the screening length and $D$ is the
ionic diffusion constant.  We propose and analyze ICEO microfluidic pumps 
and mixers that operate without moving parts under low applied 
potentials.  Similar flows around metallic colloids with fixed total charge have been 
described in the Russian literature (largely unnoticed in the West).  
ICEO flows around conductors with fixed potential, on the other hand, have no 
colloidal analog and offer further possibilities for microfluidic
applications.
\end{abstract}

\section{Introduction}
\label{sec:intro}
\hyphenation{elec-tro-pho-retic} 

Recent developments in micro-fabrication and the technological promise
of microfluidic `labs on a chip' have brought a renewed interest to
the study of low-Reynolds number flows
(\cite{stone01,whitesides01b,reyes02a}).  The familiar techniques used
in larger-scale applications for fluid manipulation, which often
exploit fluid instabilities due to inertial nonlinearities, do not
work on the micron scale due to the pre-eminence of viscous damping.
The microscale mixing of miscible fluids must thus occur without the
benefit of turbulence, by molecular diffusion alone.  For extremely
small devices, molecular diffusion is relatively rapid; however, in
typical microfluidic devices with 10-100 $\mu$m features, the mixing
time can be prohibitively long (of order 100 s for molecules with
diffusivity $10^{-10}$ m$^2$/s).  Another limitation arises because
the pressure-driven flow rate through small channels decreases with
the third or fourth power of channel size.  Innovative ideas are thus
being considered for pumping, mixing, manipulating and separating on
the micron length scale (\eg \cite{beebe02,whitesides01a}).
Naturally, many focus on the use of surface phenomena, owing to the
large surface to volume ratios of typical microfluidic devices.

Electrokinetic phenomena provide one of the most popular
non-mechanical techniques in microfluidics.  The basic idea behind
electrokinetic phenomena is as follows: locally non-neutral fluid
occurs adjacent to charged solid surfaces, where a diffuse cloud of
oppositely-charged counter-ions `screen' the surface charge.  An
externally-applied electric field exerts a force on this charged
diffuse layer, which gives rise to a fluid flow relative to the
particle or surface.  Electrokinetic flow around stationary surfaces
is known as electro-osmotic flow, and the electrokinetic motion of
freely-suspended particles is known as electrophoresis.
Electro-osmosis and electrophoresis find wide application in
analytical chemistry (\cite{bruin00}), genomics (\cite{landers03}) and
proteomics (\cite{figeys01,dolnik01}).

The standard picture for electrokinetic phenomena involves surfaces
with fixed, constant charge (or, equivalently, zeta potential $\zeta$,
defined as the potential drop across the screening cloud).  Recently,
variants on this picture have been explored.  \cite{anderson85b}
demonstrated that interesting and counter-intuitive effects occur with
spatially inhomogeneous zeta-potentials, and showed that the
electrophoretic mobility of a colloid was sensitive to the {\sl
distribution} of surface charge, rather than simply the total net
charge.  \cite{anderson85a} explored electro-osmotic flow in
inhomogeneously-charged pores, and found eddies and recirculation
regions.  
Ajdari (1995; 2001) and \cite{stroock00} showed that a net electro-osmotic flow could be
driven either parallel or perpendicular to an applied field by
modulating the surface and charge density of a microchannel, and
\cite{gitlin03} have implemented these ideas to make a `transverse
electrokinetic pump'.  Such transverse pumps have the advantage that a
strong field can be established with a low voltage applied across a
narrow channel.  
\cite{long98} examined electrophoresis of patterned colloids, and
found example colloids whose electrophoretic motion is always
transverse to the applied electric field.  Finally, \cite{long99},
\cite{ghosal03}, and others have studied electro-osmosis along
inhomogeneously-charged channel walls (due to, \eg, adsorption of 
analyte molecules), which provides an additional source of dispersion
that can limit resolution in capillary electrophoresis.

Other variants involve surface charge densities that are not fixed,
but rather are induced (either actively or passively).  For example,
the effective zeta potential of channel walls can be manipulated
using an auxillary electrode to improve separation efficiency in
capillary electrophoresis (\cite{lee90,hayes92}) and, by analogy with
the electronic field-effect transistor, to set up `field-effect
electro-osmosis' (\cite{ghowsi91b,gajar92,schasfoort99}).

Time-varying, inhomogeneous charge double-layers induced around
electrodes give rise to interesting effects as well.  \cite{trau97}
and \cite{yeh97} demonstrated that colloidal spheres can spontaneously
self-assemble into crystalline aggregates near electrodes under AC
applied fields. They proposed somewhat similar electrohydrodynamic
mechanisms for this aggregation, in which an inhomogeneous screening
cloud is formed by (and interacts with) the inhomogeneous applied
electric field (perturbed by the sphere), resulting in a rectified
electro-osmotic flow directed radially in toward the sphere.  More
recently, 
\cite{nadal02b} performed detailed measurements in order to
test both the attractive (electrohydrodynamic) and repulsive
(electrostatic) interactions between the spheres, and
\cite{ristenpart03} explored the rich variety of patterns that form
when bidisperse colloidal suspensions self-assemble near electrodes.

A related phenomenon allows steady electro-osmotic flows to be driven
using AC electric fields.  
Ramos {\it et al.} (1998; 1999) and \cite{gonzalez00} theoretically
and experimentally explored `AC electro-osmosis', in which a pair of
adjacent, flat electrodes located on a glass slide and subjected to AC
driving, gives rise to a steady electro-osmotic flow consisting of two
counter-rotating rolls. Around the same time, \cite{ajdari00}
theoretically predicted that an {\sl asymmetric} array of electrodes
with applied AC fields generally pumps fluid in the direction of
broken symmetry (`AC pumping').  \cite{brown01}, \cite{studer02}, and
\cite{mpholo03} have since developed AC electrokinetic micro-pumps
based on this effect, and \cite{ramos03} have furthered their analysis.

Both AC colloidal self-assembly and AC electro-osmosis occur around
electrodes whose potential is externally controlled. Both effects are
strongest when the voltage oscillates at a special AC frequency (the 
inverse of the charging time discussed below), and both effects
disappear in the DC limit.  Furthermore, both vary with the {\sl
square} of the applied voltage $V_0$.  This nonlinear dependence can
be understood qualitatively as follows: the induced charge cloud/zeta potential
varies linearly with $V_0$, and
the resulting flow is driven by the external field, which also varies
with $V_0$.  On the other hand, DC colloidal aggregation, as explored
by \cite{solomentsev97}, requires large enough voltages to pass a
Faradaic current, and is driven by a different, linear mechanism.

Very recently in microfluidics, a few cases of non-linear
electro-osmotic flows around isolated and inert (but polarizable)
objects have been reported, with both AC and DC forcing. In a
situation reminiscent of AC electro-osmosis, \cite{nadal02a} studied
the micro-flow produced around a dielectric stripe on a planar
blocking electrode. In rather different situations, \cite{thamida02}
observed a DC nonlinear electrokinetic jet directed away from a
protruding corner in a dielectric microchannel, far away from any
electrode, and \cite{takhistov03} observed electrokinetically driven
vortices near channel junctions. These studies (and the present work) suggest that a rich
variety of nonlinear electrokinetic phenomena at polarizable surfaces
remains to be exploited in microfluidic devices.

In colloidal science, nonlinear electro-osmotic flows around
polarizable (metallic or dielectric) particles were studied 
almost two decades ago in a series of Ukrainian papers, reviewed by
\cite{murtsovkin96}, that has gone all but unnoticed in the
West. Such flows occur when the applied field acts on the component of
the double-layer charge that has been polarized by the field itself.
This idea can be traced back at least to \cite{levich62}, who discussed
the dipolar charge double layer (using the Helmholtz model) that is induced around a 
metallic colloidal particle in an external electric field
and touched upon the quadrupolar flow that would result.  \cite{simonov73a}
calculated the structure of the (polarized) dipolar charge cloud 
in order to obtain the electrophoretic mobility, without concentrating
on the resulting flow.
\cite{gamayunov86} and \cite{dukhin86a} first explicitly calculated
the nonlinear electro-osmotic flow arising from double-layer
polarization around a spherical conducting particle, and
\cite{dukhin86b} extended this calculation to include a dielectric
surface coating (as a model of a dead biological cell).  Experimentally,
\cite{gamayunov92} observed a non-linear flow around spherical metallic colloids, 
albeit in a direction opposite to predicitions for all but the smallest particles.
They argued that a Faradaic current (breakdown of ideal polarizibility) 
was responsible for the observed flow reversal.

This work followed naturally from many earlier studies on
`non-equilibrium electric surface phenomena' reviewed by
\cite{dukhin93}, especially those focusing on the `induced dipole
moment' (IDM) of a colloidal particle reviewed by \cite{dukhin80}.
Following \cite{overbeek43}, who was perhaps the first to consider
non-uniform polarization of the double layer in the context of
electrophoresis, \cite{dukhin65}, \cite{dukhin70}, and \cite{dukhin74}
predicted the electrophoretic mobility of a highly charged
non-polarizable sphere in the thin double-layer limit, in good
agreement with the later numerical solutions of \cite{obrien78}. (See, e.g.,
\cite{lyklema91}.)   
In that case, diffuse charge is redistributed by
surface conduction, which produces an IDM aligned with the field and
some variations in neutral bulk concentration, and secondary
electro-osmotic and diffusio-osmotic flows develop as a result.
\cite{shilov70} extended this work to a non-polarizable sphere in an
AC electric field.  
\cite{simonov73a} and \cite{simonov73b} performed 
similar calculations for an ideally polarizable, conducting sphere, which typically exhibits
an IDM opposite to the applied field.  \cite{simonov77a} revisited
this problem in the context of dielectric dispersion and proposed a
much simpler RC-circuit model to explain the sign and frequency
dependence of the IDM.   
\cite{obrien78} performed a numerical solution
of the full ion, electrostatic, and fluid equations with arbitrarily
thick double-layer, which naturally incorporated the effects of double-layer
polarization.   \cite{obrien81} and \cite{obrien83}, following Dukhin's 
approach, arrived at a simpler, approximate expression that incorporated 
double-layer polarization in the thin double-layer limit, and compared favorably with the
numerical calculations of \cite{obrien78}.  Nonetheless, it seems a
detailed study of the associated electro-osmotic flows around polarizable
spheres did not appear until the paper of \cite{gamayunov86} cited above.

In summary, we note that electrokinetic phenomena at polarizable
surfaces share a fundamental feature: all involve a nonlinear flow
component in which double-layer charge {\sl induced} by the applied
field is driven by that same field.  To emphasize this common
mechanism, we suggest the term `induced-charge electro-osmosis' (ICEO)
for their description. Specific realizations of ICEO include AC
electro-osmosis at micro-electrodes, AC pumping by asymmetric
electrode arrays, DC electrokinetic jets around dielectric structures,
DC and AC flows around polarizable colloidal particles, and the
situations described below. Of course, other electrokinetic effects
may also occur in addition to ICEO in any given system, such as those
related to bulk concentration gradients produced by surface conduction
or Faradaic reactions, but we ignore such complications here to
highlight the basic effect of ICEO.

In the present work, we build upon this foundation of induced-charge
electrokinetic phenomena, with a specific eye towards microfluidic
applications. ICEO flows around metallic colloids, which have 
attraced little attention compared to
non-polarizable objects of fixed zeta potential, naturally lend
themselves for use in microfluidic devices. In that setting, the
particle is replaced by a fixed polarizable object which pumps the
fluid in response to applied fields, and a host of new possibilities
arise. The ability to directly control the position, shape, and
potential of one or more `inducing surfaces' in a microchannel allows
for a rich variety of effects that do not occur in colloidal
systems. 

Before we begin, we note the difference between ICEO and
`electrokinetic phenomena of the second kind', reviewed by
\cite{dukhin91} and studied recently by \cite{ben02} in the context of
microfluidic applications.  Significantly, second-kind
electrokinetic effects do {\sl not} arise from the double-layer, being
instead driven by space charge in the bulk solution.  They typically
occur around ion-selective porous granules subject to applied fields
large enough to generate strong currents of certain ions through the
liquid/solid interface. This leads to large concentration variations
and space charge in the bulk electrolyte on one side, which interact
with the applied field to cause motion. \cite{barany98} studied the
analogous effect for non-porous metallic colloids undergoing
electrochemical reactions at very large Faradaic currents (exceeding
diffusion limitation).  In contrast, ICEO occurs around 
inert polarizable surfaces carrying no Faradaic current in contact
with a homogeneous, quasi-neutral electrolyte and relies on relatively
small induced double-layer charge, rather than bulk space charge.

The article is organized as follows: \S\ref{sec:background}
provides a basic background on electrokinetic effects, and
\S\ref{sec:picture} develops a basic physical picture of induced-charge
electro-osmosis via calculations of steady ICEO flow around a
conducting cylinder.  \S\ref{sec:unsteady} examines the
time-dependent ICEO flow for background electric fields which are
suddenly-applied (\S\ref{sec:sudden}) or sinusoidal (\S\ref{sec:ACiceo}).
\S\ref{sec:microf} describes some basic issues for ICEO in
microfluidic devices, such as coupling to the external circuit (\S\ref{sec:electrodes}) and the
phenomenon of fixed-potential ICEO (\S\ref{sec:fixedpot}). Some specific designs for
microfluidic pumps, junction switches, and mixers are discussed and
analyzed in \S{\ref{sec:devices}. \S\ref{sec:dielectric}
investigates the detrimental effect of a thin dielectric coating on a
conducting surface and calculates the ICEO flow around non-conducting
dielectric cylinders. \S\ref{sec:rigorous} gives a systematic
derivation of ICEO in the limit of thin double layers and small
potentials, starting with the basic electrokinetic equations and
employing matched asymptotic expansions, concluding with a set of
effective equations (with approximations and errors quantified) for
the ICEO flow around an arbitrarily-shaped particle in an arbitrary
space- and time-dependent electric field.  The interesting
consequences of shape and field asymmetries, which generally lead to
electro-osmotic pumping or electrophoretic motion in AC fields, are
left for a companion paper.  The reader is referred to
\cite{bazant04a} for an overview of our results.

\section{Classical (`fixed-charge') electro-osmosis}
\label{sec:background}

Electrokinetic techniques provide some of the most popular small-scale
non-mechanical strategies for manipulating particles and fluids.  We
present here a very brief introduction.  More detailed accounts are
given by \cite{lyklema91}, \cite{hunter00} and \cite{russel89}.

\begin{figure}
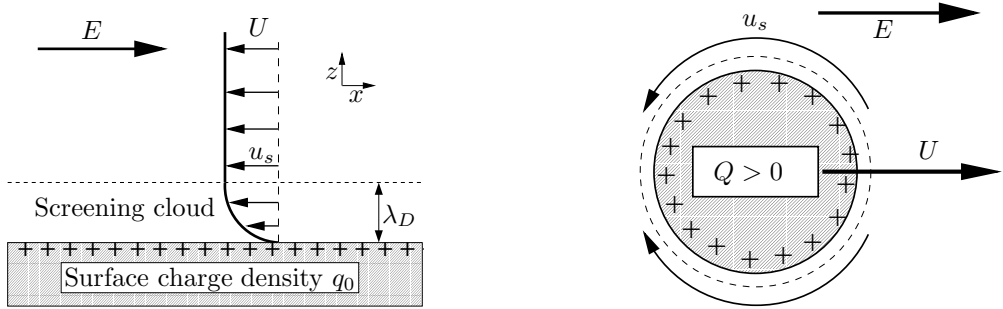

\begin{center}
\centerline{
\input{electroosmosis.pstex_t}\hfill
\input{electrophoresis.pstex_t}}
\caption{\label{fig:electrophoresis} Left: A charged solid surface in an
electrolytic solution attracts an oppositely-charged `screening cloud' of
width $\sim \lambda_D$.  An electric field applied tangent to the charged
solid surface gives rise to an electro-osmotic flow (\ref{eq:smoluchowski}).  Right: An electric field applied to an
electrolytic solution containing a suspended solid particle gives rise to
particle motion called electrophoresis, with velocity equal in magnitude
and opposite in direction to  (\ref{eq:smoluchowski}).}
\end{center}
\end{figure}

\subsection{ Small zeta potentials }

A surface with charge density $q$ in an aqueous solutions attracts a
screening cloud of oppositely-charged counter-ions to form the
electrochemical `double layer', which is effectively a surface capacitor. In
the Debye-H\"uckel limit of small surface charge, the 
excess diffuse ionic charge exponentially screens the electric field
set up by the surface charge (figure \ref{fig:electrophoresis}a),
giving an electrostatic potential
\be
\phi = \frac{q}{\eps \kappa}e^{-\kappa z}\equiv \zeta e^{-\kappa
z}.   \label{eq:linpot}
\ee
Here $\eps\approx 80\epz$ is the dielectric permittivity of the
solvent (typically water) and $\epz$ is the 
vacuum permittivity.  The `zeta potential', defined by
\be
\zeta \equiv \frac{q}{\eps \kappa},
\label{eq:zetacharge}
\ee
reflects the electrostatic potential drop
across the screening cloud, and the Debye `screening length' 
$\kappa^{-1}$ is defined for a symmetric $z$:$z$ electrolyte
by
\be
\kappa^{-1}\equiv \lambda_D= \sqrt{\frac{\eps k_B T}{2 n_0 (ze)^2}},
\label{eq:screeninglength}
\ee 
with bulk ion concentration $n_0$, (monovalent) ion charge $e$,
Boltzmann constant $k_B$ and temperature $T$. Because the ions in the
diffuse part of the double layer are approximately in thermal
equilibrium, the condition for a `small' charge density (or zeta
potential) is $\zeta \ll k_BT/ze$.

An externally applied electric field exerts a body force on the
electrically charged fluid in this screening cloud, driving the ions
and the fluid into motion.  The resulting {\sl electro-osmotic} fluid
flow (figure \ref{fig:electrophoresis}a) appears to `slip' just outside
the screening layer of width $\lambda_D$. Under a wide range of
conditions, the local slip velocity is given by the
Helmholtz-Smoluchowski formula,
\be
\bu_s = -\frac{\eps \zeta}{\eta}\bE_{||} ,
\label{eq:smoluchowski}
\ee where $\eta$ is the fluid viscosity and $\bE_{||}$ is the
tangential component of the bulk electric field. 

This basic electrokinetic phenomenon gives rise to two related effects,
electro-osmosis and electrophoresis, both of which find wide application in
analytical chemistry, microfluidics, colloidal self-assembly, and other
emerging technologies.  Electro-osmotic flow occurs when an electric field
is applied along a channel with charged walls, wherein the electro-osmotic
slip at the channel walls gives rise to plug flow in the channel, with
velocity given by (\ref{eq:smoluchowski}).  Because the
electro-osmotic flow velocity is independent of channel size, (in contrast
to pressure-driven flow, which depends strongly upon channel size),
electro-osmotic pumping presents a natural and popular technique for fluid
manipulation in small channels.
On the other hand, when the  solid/fluid interface is that of a
freely-suspended particle, the electro-osmotic slip velocity gives rise to
motion of the particle itself (figure \ref{fig:electrophoresis}b), termed
{\sl electrophoresis}. In the thin double-layer limit, the electrophoretic
velocity is given by Smoluchowski's formula, 
\be
\bU = \frac{\eps \zeta}{\eta}\bE_\infty \equiv \mu_e \bE_\infty, \label{eq:emob}
\ee
where $\bE_\infty$ is the externally-applied field, and $\mu_e = \eps
\zeta/\eta$ is the electrophoretic mobility of the particle.

\subsection{ Large zeta potentials }

For `large' zeta potentials, comparable to or exceeding $k_BT/ze$, the
exponential profile of the diffuse charge (\ref{eq:linpot}) and the
linear charge-voltage relation (\ref{eq:zetacharge}) are no longer
valid, but the diffuse part of the double layer remains in thermal
equilibrium.  As a result, the potential in the
diffuse layer satisfies the Poisson-Boltzmann equation. For symmetric,
binary electrolyte, the classical Gouy-Chapman solution yields a
nonlinear charge-voltage relation for the double layer,
\be
q(\zeta) = 4 n_0 ze \lambda_D\, \sinh\left(\frac{ze\zeta}{2 k_B T}\right),
 \label{eq:q_GC}
\ee
in the thin double-layer limit.
This relation may be modified to account for microscopic surface
properties, such as a compact Stern layer, a thin dielectric coating,
or Faradaic reactions, which enter via the boundary conditions to the
Poisson-Boltzmann equation. The key point here is simply that $q$
generally grows exponentially with $ze\zeta/k_B T$, which has important
implications for time-dependent problems involving double-layer
relaxation, as reviewed by \cite{bazant04b}.

Remarkably, the Helmholtz-Smoluchowski formula (\ref{eq:smoluchowski})
for the electro-osmotic slip remains valid in the nonlinear regime, as
long as
\be
\frac{\lambda_D}{a} \, \exp\left(\frac{ze\zeta}{2k_B T}\right) \ll 1,
\label{eq:valid}
\ee
where $a$ is the radius of curvature of the surface (Hunter 2000). 
When (\ref{eq:valid}) is violated, ionic concentrations in
the diffuse layer differ significantly from their bulk values, and
surface conduction through the diffuse layer becomes important. As a
result, the electrophoretic mobility, $\mu_e$, becomes a nonlinear
function of the dimensionless ratio 
\be
\Du = \frac{\sigma_s}{\sigma a},   
\label{eq:Dudef}
\ee
of the surface conductivity, $\sigma_s$, to the bulk conductivity,
$\sigma$.  Though this ratio was first noted by \cite{bikerman40},  
we follow \cite{lyklema91} in referring to (\ref{eq:Dudef}) as the `Dukhin number', in
honor of Dukhin's pioneering calculations of its effect on
electrophoresis.  We note also that essentially the same number has
been called $\alpha$, $\beta$, and $\lambda$ by various authors in
the Western literature, and `${\rm Rel}$' by \cite{dukhin93}. 

Bikerman (1933; 1935) made the first theoretical predictions of
surface conductance, $\sigma_s$, taking into account both
electromigration and electro-osmosis in the diffuse layer. The 
relative importance of the latter is determined by another
dimensionless number
\be
m = \left(\frac{k_B T}{ze}\right)^2 \frac{2\varepsilon_w}{\eta D}.
\ee
Using the result of \cite{deryagin69}, Bikerman's formula for
$\sigma_s$ takes a simple form for a symmetric binary electrolyte, 
yielding
\be
\Du = \frac{2\lambda_D(1+m)}{a} \left[ \cosh\left(
\frac{ze\zeta}{2k_B T}\right) - 1 \right] = 4\frac{\lambda_D(1+m)}{a} \sinh^2\left(
\frac{ze\zeta}{4k_B T}\right)
   \label{eq:Du}
\ee
for the Dukhin number. In the limit of infinitely thin double layers, where
(\ref{eq:valid}) holds, the Dukhin number vanishes, and the
electrophoretic mobility tends to Smoluchowski's value
(\ref{eq:smoluchowski}). For any finite double-layer thickness,
however, highly charged particles (with $\zeta > k_B T/ze$) are generally
described by a non-negligible Dukhin number, and surface conduction
becomes important. This leads to bulk concentration gradients and a
non-uniform diffuse-charge distribution around the particle in an
applied field, which modifies its electrophoretic mobility, via
diffusiophoresis and concentration polarization. For more details,
the reader is referred to \cite{lyklema91} and \cite{dukhin93}.

\section{Induced-charge electro-osmosis: fundamental picture }
\label{sec:picture}

The standard electrokinetic picture described above involves the interaction
of an applied field and a surface of fixed charge, wherein the electro-osmotic flow is linear in the applied
field.  Here we focus on the {\it nonlinear} phenomenon of ICEO
at a {\it polarizable} (metal or dielectric) surface.  As a
consequence of nonlinearity, ICEO can be used to drive steady
electro-osmotic flows using AC or DC fields.  The
nonlinearity also allows for larger fluid velocities and a richer,
geometry-dependent flow structure. These properties stand in stark
contrast to classical electro-osmosis described above, which, \eg,
gives zero time-averaged flow in an AC field.

This section gives a physically clear (as well as quantitatively accurate) sense of
ICEO in perhaps the simplest case: an ideally polarizable metal
cylinder in a suddenly applied, uniform electric field. This builds on
the descriptions of double-layer polarization for an uncharged
metallic sphere by \cite{levich62}, \cite{simonov73b} and
\cite{simonov77a} and the associated steady ICEO flow by
\cite{gamayunov86}.  We postpone for \S\ref{sec:rigorous} a
more detailed and general analysis, justifying the approximations made
here for the case of thin double layers ($\lambda_D \ll a$) and weak
applied fields ($E_0 a \ll k_B T/ze$). Since $\Du\ll 1$ in these limits, surface
conduction and concentration polarization can be safely ignored, and ICEO 
becomes the dominant electrokinetic effect around any inert, highly polarizable object.

\begin{figure}
\begin{center}
\centerline{
\includegraphics[width=2.5in]{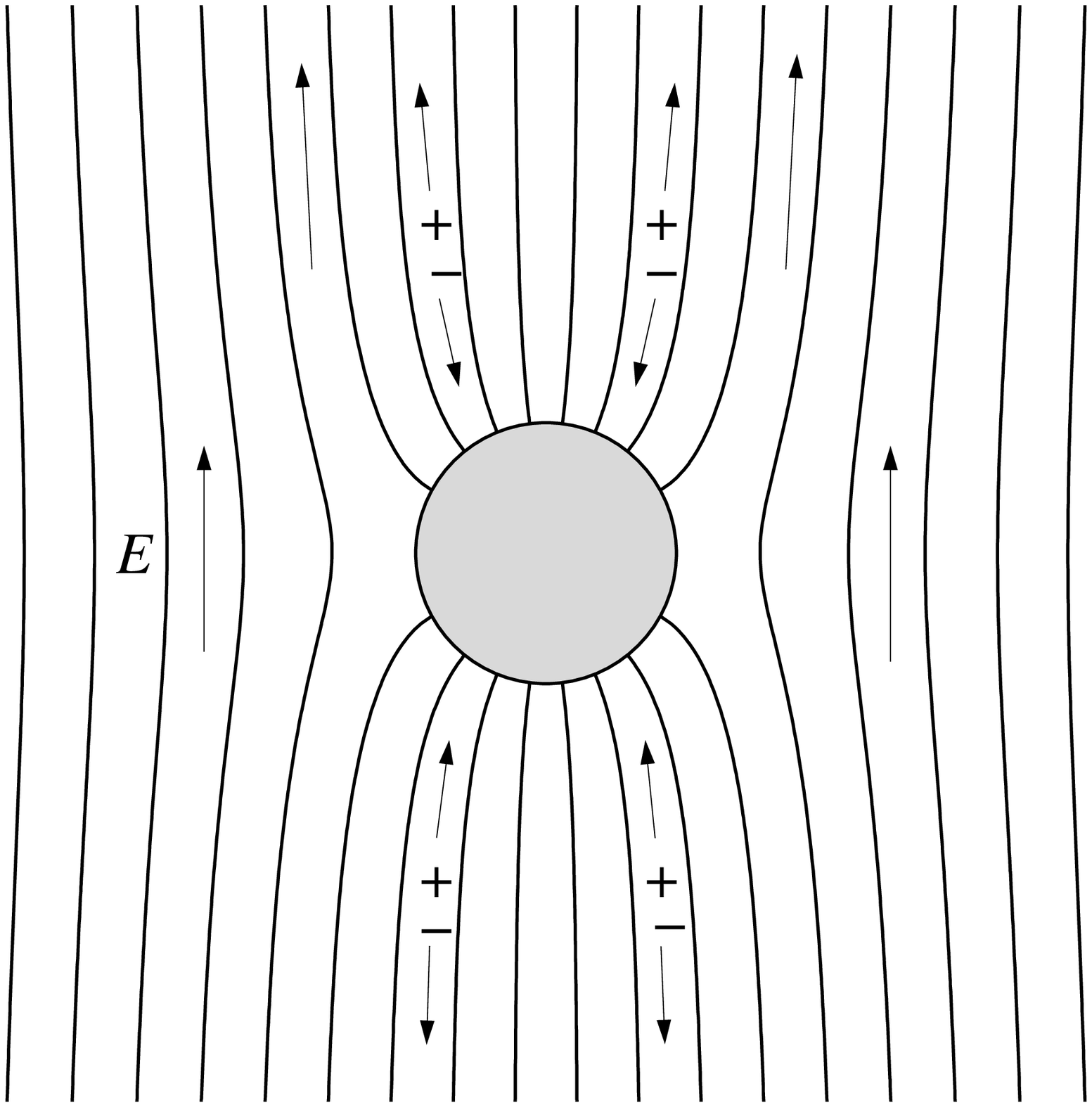}\hfill
\includegraphics[width=2.5in]{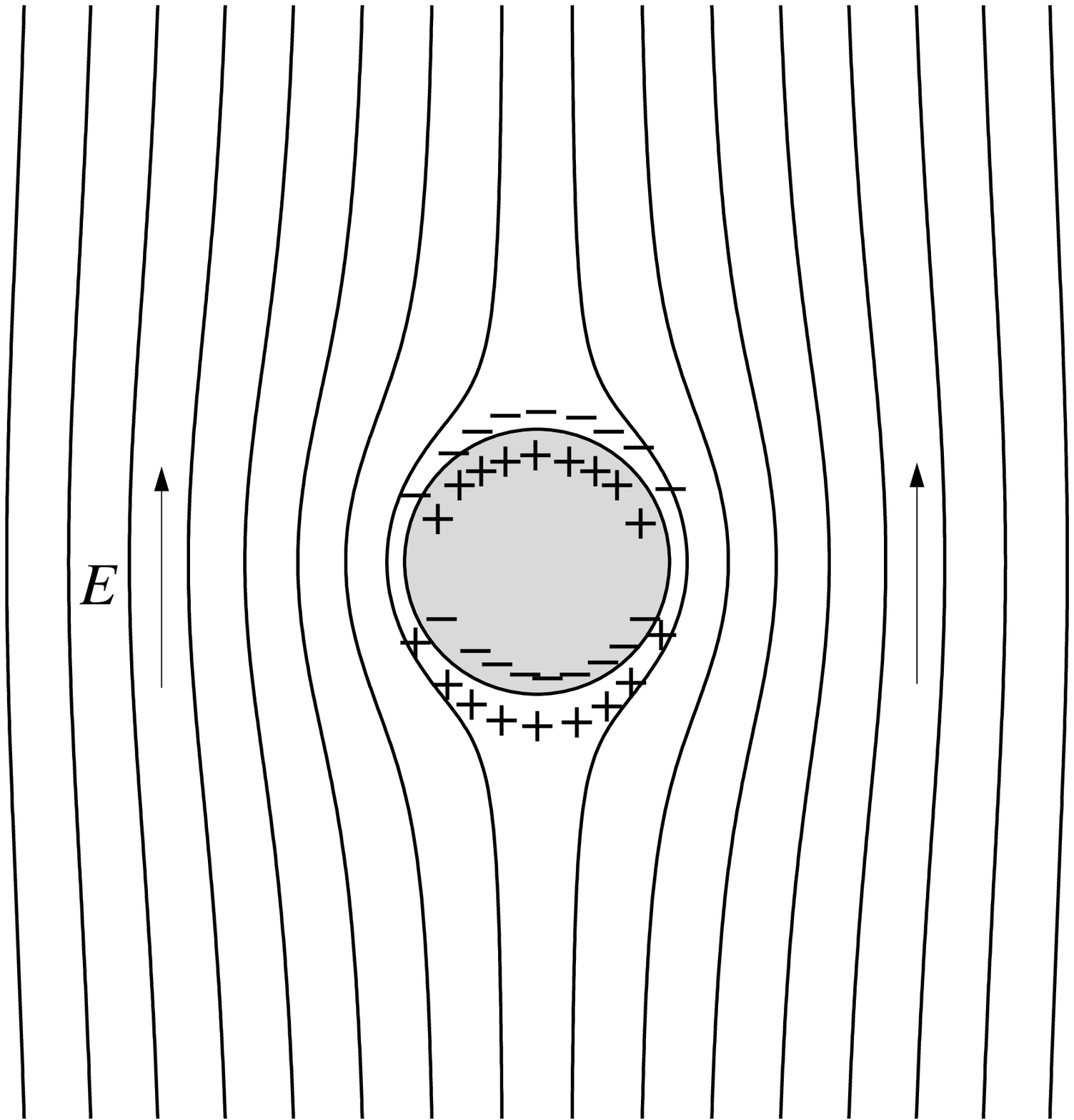}}
\caption{\label{fig:initialfinalfields} The evolution of the electric
field around a solid, ideally polarizable conducting cylinder immersed
in a liquid electrolyte, following the imposition of a background DC
field at $t=0$ (a), where the field lines intersect normal to the
conducting surface.  Over a charging time $\tau_c = \lambda_D a/D$,
a dipolar charge cloud forms in response to currents from the bulk,
reaching steady state (b) when the bulk field profile is that of an
insulator. The resulting zeta potential, however, is nonuniform.}
\end{center}
\end{figure}

\begin{figure}
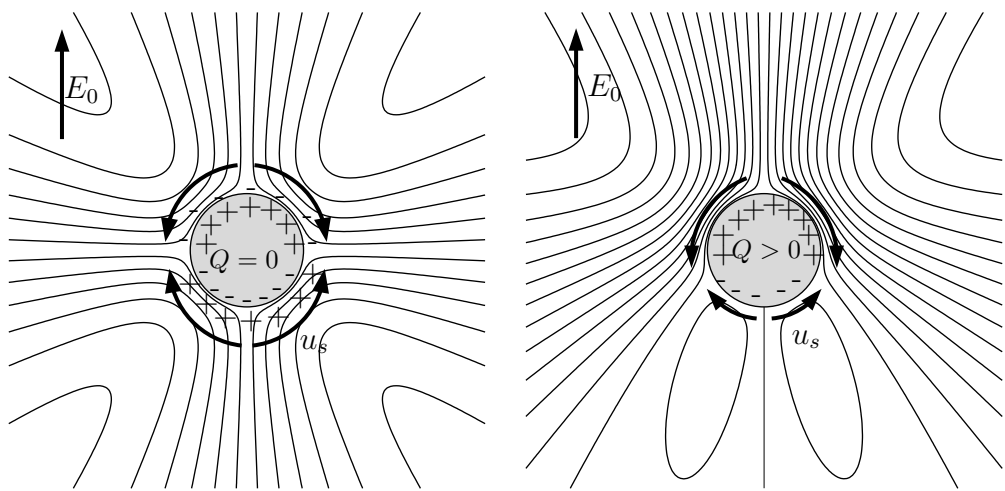

\begin{center}
\centerline{
\input{cylinderflow.pstex_t}\hfill
\input{chargedcylflow.pstex_t}}
\caption{\label{fig:cylinderflow} The steady-state induced-charge
electro-osmotic flow around (a) a conducting cylinder with zero net
charge and (b) a positively charged conducting cylinder.  The ICEO slip
velocity depends on the product of the steady field and the induced
zeta potential.  The flow around an uncharged conducting cylinder (a) 
can thus be understood qualitatively from figure \ref{fig:initialfinalfields}b,
whereas the charged sylinder (b) simply involves the superposition of the
standard electro-osmotic flow.}
\end{center}
\end{figure}

\subsection{Steady ICEO around an uncharged conducting cylinder}
\label{sec:steady}

The basic phenomenon of ICEO can be understood from figures
\ref{fig:initialfinalfields} and \ref{fig:cylinderflow}.  Immediately
after an external field $\bE=E_0 \zh$ is applied, an electric
field is set up so that field lines intersect conducting surfaces at
right angles (figure \ref{fig:initialfinalfields}a).  Although this
represents the steady-state {\sl vacuum} field configuration, mobile ions in
electrolytic  
solutions move in response to applied fields. A current $\bJ =
\sigma \bE$ drives positive ions into a charge cloud on one side of
the conductor ($z<0$), and negative ions to the other ($z>0$), inducing
an equal and opposite surface charge on the conducting surface.  
A dipolar charge cloud grows so long as a normal field injects ions into
the induced double layer, and steady state is achieved when 
no field lines penetrate the double-layer
(figure \ref{fig:initialfinalfields}b).  The
tangential field $E_{||}$ drives an electro-osmotic slip velocity
(\ref{eq:smoluchowski}) proportional to the local double-layer
charge density, driving fluid from the `poles' of the particle towards
the `equator' (figure \ref{fig:cylinderflow}a).  An AC field drives an
identical flow, since an oppositely-directed field induces an oppositely-charged screening
cloud, giving the same net flow.
The ICEO flow around a conducting cylinder with non-zero total charge, shown in figure \ref{fig:cylinderflow}b,
simply superimposes the nonlinear ICEO flow of figure \ref{fig:cylinderflow}a with the usual linear electro-osmotic flow.
 
As a concrete example for quantitative analysis, we consider an isolated, uncharged conducting cylinder of
radius $a$ submerged in an electrolyte solution with very small
screening length $\lambda_D\ll a$.  An external electric field $E_0
\zh$ is suddenly applied at $t=0$, and the conducting surface forms 
an equipotential surface, giving a potential 
\be \phi_0 = -E_0 z\left(1
-\frac{a^2}{r^2}\right).
\label{eq:initialfield}
\ee
Electric field lines  intersect the conducting surface at right angles, as
shown in figure  \ref{fig:initialfinalfields}a.

Due to the electrolyte's conductivity $\sigma$, a non-zero
current $\bJ = \sigma \bE$ drives ions to the cylinder surface.  In the absence of
electrochemical reactions at the conductor/electrolyte interface (\ie at
sufficiently low potentials that the cylinder is `ideally
polarizable'), mobile solute ions accumulate in a screening cloud
adjacent to the solid/liquid surface, attracting equal and opposite
`image charges' within the conductor itself.  Thus the
conductor's surface charge density $q$ --- induced by the growing
screening cloud --- changes in a time-dependent fashion, via 
\be
\frac{dq(\theta)}{dt} = \bj \cdot \brh = \sigma \bE \cdot \brh.
\label{eq:charging}
\ee
Using the linear relationship (\ref{eq:zetacharge}) between
surface charge density and zeta potential, this can be expressed as
\be
\frac{d\zeta(\theta)}{dt} = \frac{\sigma }{\eps \kappa}\bE \cdot \brh.
\label{eq:dzetadt}
\ee
A dipolar charge cloud grows, since positively-charged ions are driven into
the charge cloud on the side of the conductor nearest the field source
($z<0$ in this case), and negatively-charged ions are driven into the
charge cloud on the opposite side. As ions are driven into the screening
charge cloud, field lines are expelled and the ionic flux into the charge cloud
is reduced.

The system reaches a steady state configuration when all field lines are
expelled ($\brh \cdot \bE(a)=0$). This occurs when the electrostatic
potential outside of the charge cloud is given by
\be
\phi_f = -E_0 z\left(1 +\frac{a^2}{r^2}\right),
\label{eq:finalfield}
\ee
shown in figure \ref{fig:initialfinalfields}b. The steady-state
electrostatic configuration is thus equivalent to the no-flux electrostatic
boundary condition assumed in the analysis of `standard' electrophoresis.
In the present case, however, the steady-state configuration corresponds to
a cylinder whose zeta potential varies with position according to
\be
\zeta(\theta) =\phi_o - \phi_f(a)= 2 E_0 a \cos\theta,
 \label{eq:localzeta}
\ee
where we assume the conductor potential $\phi_o$ to vanish. While
the steady-state electric field has no component normal to the charge
cloud, its tangential component,
\be
{\bf {\hat \theta}} \cdot \bE =-2 E_0 \sin\theta,
\label{eq:steadytangentialfield}
\ee
drives an induced-charge electro-osmotic flow, with
slip velocity given by (\ref{eq:smoluchowski}).  Now, however, the
(spatially-varying) surface potential $\zeta$ is given by (\ref{eq:localzeta}).  Because the charge cloud is itself dipolar,  the
tangential field drives the two sides of the charge cloud in opposite
directions---each side away from the poles---resulting in a quadrupolar
electro-osmotic slip velocity
\be
\bu_s =2 U_0 \sin2\theta\,\, {\bf {\hat \theta}},
\label{eq:cylinderslip}
\ee
where $U_0$ is the natural velocity scale for ICEO,
\be
U_0=\frac{\eps  E_0^2 a}{\eta}.
\label{eq:natscale}
\ee
One power of $E_0$ sets up the `induced-charge' screening cloud, and the second
drives the resultant electro-osmotic flow.

The fluid motion in this problem is reminiscent of that around a fluid
drop of one conductivity immersed in a fluid of another conductivity
subjected to an external electric field, studied by \cite{taylor66}.
By analogy, we find the radial and azimuthal fluid velocity components
of the fluid flow outside of the cylinder to be 
\be u_r =2\frac{a
(a^2-r^2)}{r^3}U_0 \cos 2 \theta,\,\,\,\,\,\, u_\theta
= 2\frac{a^3}{r^3}U_0 \sin 2 \theta\label{eq:ICEOu}.  
\ee 
For comparison, analogous results for the steady-state
ICEO flow around a sphere, some of which were derived by \cite{gamayunov86}, 
are given in Table \ref{tab:ICEOcylsphere}.

\begin{table}
\begin{center}
\begin{tabular}[t]{|l|c|c|}
\hline
&Cylinder&Sphere\\
\hline
&&\\
Initial potential $\phi_i$&$-E_0 z \left(1-\frac{a^2}{r^2}\right)$&$-E_0 z \left(1-\frac{a^3}{r^3}\right)$\\
&&\\
Steady-state potential $\phi_s$&$-E_0 z \left(1+\frac{a^2}{r^2}\right)$&$-E_0 z \left(1+\frac{a^3}{2 r^3}\right)$\\
&&\\
Steady-state zeta potential $\zeta$&$2 E_0 a \cos\theta$&$\frac{3}{2} E_0a\cos\theta$\\
&&\\
Radial flow $u_r$&$\frac{2a (a^2-r^2)}{r^3}U_0 \cos 2 \theta$&$\frac{9a^2 (a^2-r^2)}{16r^4}U_0 \left(1+3\cos 2 \theta\right)$\\
&&\\
Azimuthal flow $u_\theta$&$\frac{2a^3}{r^3}U_0  \sin 2 \theta$&$\frac{9a^4}{8r^4}U_0  \sin 2 \theta$\\
&&\\
Charging Timescale  &  $\tau_c = \frac{\lambda_D a}{D}$   &   $\tau_s = \frac{\lambda_D a}{D}$\\
&&\\
Induced dipole strength:  &&\\
   $g(t)$ for suddenly applied field $\bE_0$   & $g(t) = 1-2 e^{-t/\tau_c}$ & $g(t) = \frac{1}{2}\left(1-3 e^{-2t/\tau_s}\right)$ \\
&&\\
   $g$ for AC field $\Real(\bE_0 e^{i \omega t})$                              & $g = \frac{1-i \omega \tau_c}{1+i \omega \tau_c}$ & $g = \frac{1-i \omega \tau_s}{2+i \omega \tau_s}$ \\
&&\\
\hline
\end{tabular}
\caption{Electrostatic and hydrodynamic quantities for the induced-charge
electro-osmotic (ICEO) flow around conducting spheres and cylinders, each
of radius $a$.  Here $U_0 = \eps E_0^2 a/\eta$ is a characteristic
velocity scale, and the induced dipole strength $g$ is defined in (\ref{eq:suddenphi}).  
See \cite{gamayunov86} for flows around metal colloidal spheres in steady and 
AC fields.}
\label{tab:ICEOcylsphere}
\end{center}
\end{table}

Although we focus on the limit of linear screening in this paper,
(\ref{eq:ICEOu}) should also hold for 
non-linear screening ($\zeta \approx k_B T/ze$) in the limit of thin
double layers, as long as the Dukhin number remains small and
(\ref{eq:valid}) is satisfied.  The relevant zeta
potential in these conditions, however, is not the equilibrium zeta
potential ($\zeta_0=0$) but the typical {\it induced} zeta potential,
$\zeta \approx E_0a$, which is roughly the applied voltage across the
particle.

Finally, although we have specifically considered a conducting cylinder, a similar
picture clearly holds for more general shapes. More generally, ICEO slip velocities around
arbitrarily-shaped inert objects in uniform applied fields are
directed along the surface from the `poles' of the object (as defined
by the applied field), towards the object's `equator'. 

\subsection{ Steady ICEO around a charged conducting cylinder }

Until now, we have assumed the cylinder to have zero net charge for
simplicity.  A cylinder with non-zero equilibrium charge density
$q_0=Q/4\pi a^2$, or zeta potential,
$\zeta_0 = \eps \kappa q_0$, in a suddenly applied field approaches a
steady-state zeta-potential distribution,
\be
\zeta(\theta) = \zeta_0 + 2 E_0 a \cos \theta ,  \label{eq:zetatotal}
\ee
which has the same induced component in (\ref{eq:localzeta}) 
added to the constant equilibrium value. This follows from the
linearity of (\ref{eq:dzetadt}) with the initial condition,
$\zeta(\theta,t=0)=\zeta_0$. As a result, the steady-state
electro-osmotic slip is simply a superposition of the `standard'
electro-osmotic flow due to the equilibrium zeta potential $\zeta_0$,
\be 
\bu_s^Q=\bu_s -2 \frac{\eps \zeta_0}{\eta}\sin\theta \,\,\thh,
\label{eq:standardzeta}
\ee 
where $\bu_s$ is the ICEO slip velocity, given in (\ref{eq:cylinderslip}).  The associated Stokes flow, which is a
superposition of the ICEO flow and the `standard' electro-osmotic
flow, and is shown in figure \ref{fig:cylinderflow}(b).

The electrophoretic velocity of a charged conducting cylinder can be
found using the results of \cite{stone96}, from which it follows
that the velocity of a cylinder with prescribed slip velocity
$\bu_s(\theta)$, but no externally-applied force, is given by the
surface-averaged velocity,  
\be 
\bU = -\frac{1}{2 \pi}\int_0^{2\pi} \bu_s(\theta) d\theta.  
\ee 
The ICEO component (\ref{eq:cylinderslip}) has zero surface average, 
leaving only the `standard' electro-osmotic slip velocity (\ref{eq:standardzeta}).  
This was pointed out by \cite{levich62} using a Helmholtz model for the induced
double layer, and later by \cite{simonov73a} using a double-layer structure found
by solving the electrokinetic equations.  The conducting cylinder thus has
the same electrophoretic mobility $\mu_e=\eps\zeta_0/\eta$ as an object of fixed
uniform charge density and constant zeta potential.  This extreme case
illustrates the result of \cite{obrien78} that the electrophoretic mobility
does not depend on electrostatic boundary
conditions, even though the flow around the particle clearly does.

As above, the steady-state analysis of ICEO for a charged conductor is
unaffected by nonlinear screening, as long as (\ref{eq:valid})
is satisfied (and $\Du \ll 1$), where the relevant $\zeta$ is the
maximum value, $\zeta = |\zeta_0| + |E_0a|$, including both equilibrium
and induced components.

\section{Time-dependent ICEO}
\label{sec:unsteady}

A significant feature of ICEO flow is its dependence on the square of
the electric field amplitude. This has important consequences for AC
fields: if the direction of the electric field in the above picture is
reversed, so are the signs of the induced surface charge and screening
cloud.  The resultant ICEO flow, however, remains unchanged: the net
flow generically occurs away from the poles, and towards the equator.
Therefore, induced-charge electro-osmotic flows persist even in an AC
applied fields, so long as the frequency is sufficiently low that the
induced-charge screening clouds have time to form.

AC forcing is desirable in microfluidic devices, so
it is important to examine the time-dependence of ICEO flows.  As
above, we explicitly consider a conducting (ideally polarizable)
cylinder and simply quote the analogous results for a conducting
sphere in Table \ref{tab:ICEOcylsphere}.  Although we perform
calculations for the more tractable case of linear screening, we
briefly indicate how the analysis would change for large induced zeta
potentials.  Two situations of interest are presented: the
time-dependent response of a conducting cylinder to a
suddenly-applied electric field (\S\ref{sec:sudden}) and a sinusoidal AC electric
field (\S\ref{sec:ACiceo}). We also comment on the basic time scale for ICEO flows.

\subsection{ICEO around a conducting cylinder in a suddenly-applied DC
  field}
\label{sec:sudden}

Consider first the time-dependent response of an uncharged conducting
cylinder in an electrolyte when a uniform electric field $\bE = E_0{\bf
\zh}$ is suddenly turned on at $t=0$.  The dipolar nature of the external
driving suggests a bulk electric field of the form
\be
\phi(\br,t) = -E_0z \left(1+g(t) \frac{a^2}{r^2}\right),
\label{eq:suddenphi}
\ee
so that initially $g(0)=-1$ (\ref{eq:initialfield}), and
in steady state $g(t\rightarrow\infty)=1$ (\ref{eq:finalfield}).
The potential of the conducting surface itself remains zero, so
that the potential drop across the double layer is given by
\be
\phi(a,\theta,t)= -\zeta(\theta)=-\frac{q(\theta)}{\eps
\kappa}. \label{eq:zetaphi}
\ee
Here, as before, we take $q$ to represent the induced  surface charge, so
that the total charge per unit area in the charge cloud is $-q$. The
electric field normal to the surface, found from (\ref{eq:suddenphi}), drives an ionic current
\be
J_\perp = -\dot{q}(\theta)=-\sigma E_0 \cos\theta \left(1-g\right),
\label{eq:jeperp}
\ee
into the charge cloud, locally injecting a surface charge density
$\dot{q}$ per unit time. We express the induced charge density $q$ in terms
of the induced dipole $g$ by substituting (\ref{eq:suddenphi}) into
(\ref{eq:zetaphi}), take a time derivative, and equate the result with
$\dot{q}$ given by (\ref{eq:jeperp}).  This results in an ordinary
differential equation for the dipole strength $g$,
\be
\dot{g} = \frac{\sigma}{\eps \kappa a}(1-g),
\ee
whose solution is
\be
g(t) = 1-2e^{-t/\tau_c}.
\ee
Here $\tau_c$ is the characteristic time for the formation of 
induced-charge screening clouds,
\be
\tau_c = \frac{\kappa a \eps}{\sigma} =  \frac{\lambda_D a}{D}, 
\label{eq:chargingtime}
\ee
where the definitions of conductivity ($\sigma = 2 n_0 e^2 D/k_B T$)
and screening length (\ref{eq:screeninglength}) have been used. 

The induced-charge screening cloud (with equivalent zeta potential given by
(\ref{eq:zetaphi})) is driven by the tangential field (derived
from (\ref{eq:suddenphi})) in the standard way (\ref{eq:smoluchowski}), resulting in an induced-charge electro-osmotic slip
velocity
\be
\bu_s = 2 U_0 \sin 2 \theta
\left(1-e^{-t/\tau_c}\right)^2\thh. \label{eq:tdepslip}
\ee
More generally, the time-dependent slip velocity around a cylinder with a nonzero fixed charge (or equilibrium
zeta potential $\zeta_0$) can be found in similar fashion, and results in the standard ICEO slip velocity $\bu_s$ (\ref{eq:tdepslip}) with an additional term representing standard electro-osmotic slip (\ref{eq:standardzeta})
\be
\bu^Q_s = \bu_s -2 \frac{\eps \zeta_0}{\eta}\sin\theta\left(1 - e^{-t/\tau_c}\right)\thh.
\label{eq:tdepstandard}
\ee
Note that (\ref{eq:tdepstandard}) grows more quickly than (\ref{eq:tdepslip}) initially, but that ICEO slip eventually dominates in strong fields, since it varies with $E_0^2$, versus $E_0$ for the standard electro-osmotic slip.

\subsection{ICEO around a conducting cylinder in a sinusoidal AC field}
\label{sec:ACiceo}

An analogous calculation can be performed in a straightforward fashion for sinusoidal applied fields.  
Representing the electric field using complex notation, $\bE = E_0 e^{i\omega t} {\bf \zh},$
where the real part is implied, we obtain a time-dependent zeta potential
\be
\zeta = 2 E_0 a \cos \theta\, \Real\,\,\left( \frac{e^{i \omega t}}{1+i \omega \tau_c}\right),
\ee
giving an induced-charge electro-osmotic slip velocity
\be
\bu_s = 2 U_0\sin 2 \theta \left[\Real\left(\frac{e^{i \omega t}}{1+i \omega \tau_c}\right)\right]^2\thh,
\ee
with time-averaged slip velocity
\be
\langle \bu_s \rangle= \frac{U_0 \sin 2 \theta }{(1+\omega^2 \tau_c^2)}\thh.
\ee
In the low-frequency limit $\omega \tau_c \ll 1$, the double-layer fully
develops in phase with the applied field.  In the high-frequency limit
$\omega \tau_c\gg 1$, the double-layer does not have time to charge up,
attaining a maximum magnitude $(\omega \tau_c)^{-2}$ with a $\pi/2$
phase shift.  Note that this analysis assumes that the double-layer
changes quasi-steadily, which requires $\omega \ll \tau_D^{-1}.$

\subsection{ Time scales for ICEO flows }

Before continuing, it is worth emphasizing the fundamental time scales
arising in ICEO.  The basic charging time $\tau_c$ exceeds the Debye time
for diffusion across the double-layer thickness, $\tau_D = \lambda_D^2/D = \eps/\sigma$,
by a geometry-dependent factor, $a/\lambda_D$, that is typically very large. $\tau_c$ is also much
smaller than the diffusion time across the particle, $\tau_a =
a^2/D$. The appearance of this time scale for induced dipole moment
of a metallic colloidal sphere has been explained by \cite{simonov77a}
using a simple RC-circuit analogy, consistent with the detailed
analysis of \cite{simonov73b}.  The same time scale, $\tau_c$, also
arises as the inverse frequency of AC electro-osmosis (\cite{ramos99})
or AC pumping (\cite{ajdari00}) at a micro-electrode array of
characteristic length $a$, where again an RC-circuit analogy has been
invoked to explain the charging process. This simple physical picture
has been criticized by \cite{scott01}, but \cite{ramos01} and
\cite{gonzalez00} have convincingly defended its validity, as in the
earlier Russian papers on polarizable colloids.

Although it is apparently not well known in microfluidics and
colloidal science, the time scale for double-layer relaxation was
debated and analyzed in electrochemistry in the middle of the last
century, after \cite{ferry48} predicted that $\tau_D$ should be the
charging-time for the double layer at an electrode in a semi-infinite
electrochemical cell. \cite{buck69} explicitly
corrected Ferry's analysis to account for bulk conduction, which
introduces the macroscopic electrode separation $a$. The issue was definitively settled by
\cite{macdonald70} who explained the correct charging time, $\tau_c$,
as the `$RC$ time' for the double-layer capacitor, $C = \eps/\lambda_D$,
coupled to a bulk resistor, $R = a/\sigma$.  Similiar ideas
were also developed independently a decade later by \cite{kornyshev81}
in the context of solid electrolytes.

Ferry's model problem of a suddenly imposed surface charge-density in
a semi-infinite electrolyte (as opposed to a suddenly imposed voltage
or background field in a finite system) persists in recent textbooks
on colloidal science, such as \cite{hunter00} and \cite{lyklema91}, and only
the time scales $\tau_D$ and $\tau_a$ are presented as relevant for
double-layer relaxation. This is quite reasonable for
non-polarizable colloidal particles, but we stress that the
intermediate RC time scale, $\tau_c = \sqrt{\tau_D\tau_a}$, plays a
central role for polarizable objects that exhibit significant
double-layer charging.

We also mention nonlinear screening effects at large applied fields or
large total charges, where the maximum total zeta potential, $\zeta
\approx \zeta_0 + E_0 a$, exceeds $k_B T/ze$.  The analysis of this
section can be generalized to account for the `weakly nonlinear'
limit of thin double layers, where $\zeta > k_B T/ze$, but 
(\ref{eq:valid}) is still satisfied (and thus $\Du \ll 1$ as well). In
the absence of surface conduction, (\ref{eq:charging}) still
describes double-layer relaxation, but a nonlinear charge-voltage
relation, such as (\ref{eq:q_GC}) from Gouy-Chapman theory, must
be used. In that case, the time-dependent boundary condition
(\ref{eq:dzetadt}) on the cylinder is replaced by 
\be 
C_D(\zeta) \,
\frac{d\zeta}{dt} = \sigma \, \bE \cdot \brh,
\label{eq:nonlin_charge}
\ee
where 
\be
C_D(\zeta) = \frac{\eps}{\lambda_D} \,
\cosh\left(\frac{ze\zeta}{2k_B T}\right) \label{eq:C_GC}
\ee
is the nonlinear differential capacitance of the double
layer.

If the {\it induced} component of the zeta potential is large, due to
a strong applied field, $E_0a > k_B T/ze$, the charging dynamics in
(\ref{eq:nonlin_charge}) are no longer analytically tractable. Since
the differential capacitance $C_D(\zeta)$ increases with $|\zeta|$ in
any thin double-layer model, 
the `poles' of the cylinder along the applied field charge more
slowly than the sides.  However, the steady-state field is  
the same as for linear screening, so long as $\Du \ll 1$ .  

If the applied field is weak, but the {\it total charge} 
is large ($\zeta_0 > k_B T/ze$), (\ref{eq:nonlin_charge}) may be
linearized to obtain the same polarization dynamics and ICEO flow 
as for $\Du \ll 1$.  However, the `$RC$ time',
\be
\tau_c(\zeta_0) = R \, C_D = \tau_c \,
\cosh\left(\frac{ze\zeta_0}{2k_B T}\right), \label{eq:tcnonlin}
\ee
increases nonlinearly with $\zeta_0$, as shown by \cite{simonov77a}. 
The same time constant, with $\zeta_0$ replaced by $\zeta_0
+ E_0a$, also describes the late stages of relaxation in response to a
strong applied field.

Some implications for ICEO at large voltages in the `strongly
nonlinear' limit of thin double layers, where condition
(\ref{eq:valid}) is violated and $\Du \gg 1$, are discussed in
\S\ref{sec:nonlinear}.  In this regime, the simple circuit
approximation breaks down due to bulk diffusion, and secondary
relaxation occurs at the slow time scale, $\tau_a = a^2/D$.  For an
interdisciplinary review and detailed analysis of double-layer
relaxation (without surface conduction or flow), the reader is
referred to \cite{bazant04b}.

Finally, we note that the oscillatory component of ICEO flows may not
obey the quasi-steady Stokes equations, due to the finite time scale,
$\tau_v = a^2/\nu$, for the diffusion of fluid vorticity. (Here $\nu =
\eta/\rho$ is the kinematic viscosity and $\rho$ the fluid density.)
It is customary in microfluidic and colloidal systems to neglect the
unsteady term, $\rho \partial \bu/\partial t$, in the Stokes equations,
because ions diffuse more slowly than vorticity by a
factor of $D/\nu \approx 10^{-3}$. However, the 
natural time scale for the AC component of AC electro-osmotic flows is $\tau_c = \lambda_D a/D$,
so the importance of the unsteady term in the Stokes equations is
governed by the dimensionless parameter,
\be
\frac{\tau_v}{\tau_c} = \frac{a}{\lambda_D} \, \frac{D}{\nu}.
\ee
This becomes significant for sufficiently thin double layers,
$\lambda_D/a \approx 10^{-3}$.  
Therefore, in AC electro-osmosis and other ICEO phenomena with AC forcing 
at the charging frequency, $\omega_c = \tau_c^{-1}$, vorticity diffusion
confines the oscillating component of ICEO flow to an oscillatory boundary
layer of size $\sqrt{\nu \lambda_D a/D}$.  However, the steady component of 
ICEO flows is usually the most important, and obeys the steady Stokes equations.

\section{ ICEO in microfluidic devices }
\label{sec:microf}

We have thus far considered isolated conductors in background fields applied `at infinity',
as is standard in the colloidal context.  The further richness of ICEO phenomena becomes apparent 
in the context of microfluidic devices.  In this section, we consider ICEO in which the external field is applied by electrodes with finite, rather than infinite, separation.  Furthermore, microfluidic devices allow additional techniques not available for colloids:  the `inducing surface' can be held in place and its potential can be actively controlled.  This gives rise to `fixed-potential' ICEO, which is to be contrasted with the `fixed-total-charge' ICEO studied above.
Finally, we present a series of simple ICEO-based microfluidic devices that operate without moving parts in AC fields.

\subsection{Double-layer relaxation at electrodes}
\label{sec:electrodes}
As emphasized above, one must drive a current $J_0(t) =
\sigma E_0(t)$ to apply an electric field $E_0(t)$ in an
homogeneous electrolyte. The electrochemical reactions associated with 
steady Faradaic currents may cause fluid contamination by reaction
products or electrodeposits, unwanted concentration polarization, or
permanent dissolution (and thus irreversible failure) of
microelectrodes. Therefore, oscillating voltages and non-Faradaic 
displacement currents at `blocking' electrodes are preferable in
microfluidic devices.  In this case, however, one must take care that 
diffuse-layer charging at the electrodes does not
screen the field.

We examine the simplest case here, which involves a device consisting
of a thin conducting cylinder of radius $a \gg \lambda_D$ placed
between flat, inert electrodes separated by $2L \gg 2a$. The cylinder
is electrically isolated from the rest of the system, so that its
total charge is fixed.  Under a suddenly applied DC voltage, $2 V_0$, the
bulk electric field, $E_0(t)$, decays to zero as screening clouds
develop at the electrodes. For weak potentials ($V_0 \ll k_BT/e$) and
thin double-layers ($\lambda_D\ll a \ll L$), the bulk field decays
exponentially
\be
E_0(t) = \frac{V_0}{L}e^{- t/\tau_e},
\ee
with a characteristic electrode charging time
\be
\tau_e = \frac{\lambda_D L}{D} ,
\ee
analogous to the cylinder's charging time (\ref{eq:chargingtime}). 
This time-dependent field $E_0(t)$ then acts as the `applied
field at infinity' in the ICEO slip formula, (\ref{eq:tdepslip}).  

The ICEO flow around the cylinder is set into
motion exponentially over the cylinder charging time, $\tau_c = \lambda_D a/D$, but
is terminated exponentially over the (longer) electrode charging time, $\tau_e =
\lambda_D L/D$ as the bulk field is screened at the electrodes. This
interplay between two time scales --- one set by the geometry of the inducing
surface and another set by the electrode geometry --- is a common feature of
ICEO in microfluidic devices. 

This is clearly seen in the important case of AC forcing by a voltage,
$V_0 \cos(\omega t)$, in which the bulk electric field is given
by 
\be 
E_0(t) = \frac{V_0}{L}\Real \left( \frac{i \omega \tau_e }{1 +
i\omega \tau_e}e^{-i \omega t} \right).  
\ee 
Electric
fields persist in the bulk solution when the driving frequency is high
enough $(\omega \tau_e\gg1)$ that induced double-layers do not have time to
develop near the electrodes.  Induced-charge electro-osmotic flows
driven by applied AC fields can thus persist only in a certain band of
driving frequencies, $\tau_e^{-1} \leq \omega \leq \tau_c^{-1}$,
unless Faradaic reactions occur at the electrodes to maintain the bulk
field.  In AC electro-osmosis at adjacent surface electrodes (\cite{ramos99})
or AC pumping at an asymmetric electrode array (\cite{ajdari00}), the inducing 
surfaces {\sl are} the electrodes, and so the two time scales coincide to yield a
single characteristic frequency $\omega_c = 1/\tau_e$.
(Note that \cite{ramos99} and \cite{gonzalez00} use the equivalent form
$\omega \sim \sigma \lambda_D / \eps L$.)

Table \ref{tab:repvals} presents typical values for ICEO flow velocities and charging time scales for some
reasonable microfluidic parameters.  For example, an applied electric
field of strength $100$ V/cm across an electrolyte containing a 10
$\mu$m cylindrical post gives rise to an ICEO slip velocity of order 1
mm/s, with charging times $\tau_c \sim 0.1$ ms and $\tau_e = 0.1$~s.
\begin{table}
\begin{center}
\begin{tabular}[t]{llll}
\multicolumn{4}{c}{{\bf Material properties of aqueous solution}}\\
Dielectric constant&    $\eps\approx 80\epz$ &$7\times 10^{-5}$ &   g cm/V$^2$s$^2$\\ 
Viscosity&$\eta$ & $10^{-2}$ &g/cm s\\ 
Ion Diffusivity&$D_i$ & $10^{-5}$ &cm$^2$/s\\
\multicolumn{4}{c}{{\bf Experimental Parameters}}\\ 
Screening Length&$\lambda_D$ & 10 &nm\\ 
Cylinder radius&$a$ & 10 &$\mu$m\\ 
Applied field&$E_0$&100 &V/cm\\ 
Electrode Separation&$L$ & 1 &cm\\
\multicolumn{4}{c}{{\bf Characteristic Scales}}\\ 
Slip velocity&$U_0 = \eps  E_0^2 a/\eta$&0.7&mm/s\\ 
Cylinder charging time &$\tau_c$ = $\lambda_D a/D_i$&$10^{-4}$&s\\ 
Electrode charging time &$\tau_e$ = $\lambda_D L/D_i$&$10^{-1}$&s\\ 
Dimensionless Surface Potential &$\Psi$ = $e E_0 a/k_B T$&3.9&\\
Dukhin number &${\rm Du}$ &$10^{-2}$&\\
\hline
\end{tabular}
\caption{Representative values for induced-charge electro-osmosis in a
microfluidic device.} \label{tab:repvals}
\end{center}
\end{table}

\subsection{Fixed-potential ICEO}
\label{sec:fixedpot}

In the above examples, we have assumed a conducting element which is electrically
isolated from the driving electrodes, which constrains the total charge
on the conductor.  Another possibility involves
fixing the potential of the conducting surface with respect to the driving electrodes, 
which requires charge to flow onto and off
of the conductor. This ability to
directly control the induced charge allows for a wide variety of
microfluidic pumping strategies exploiting ICEO.

Perhaps the simplest example of fixed-potential ICEO involves a conducting
cylinder of radius $a$ which is held in place at a distance $h$ from the
nearest electrode (figure \ref{fig:fixedpot}a).  For simplicity, we consider 
$a \ll h$ and $h \ll L$. We take the cylinder to be held at some potential $V_c$, the nearest
electrode to be held at $V_0$, the other electrode at $V=0$, and assume the
electrode charging time $\tau_e$ to be long.  In this case, the bulk field
(unperturbed by the conducting cylinder) is given simply by $E_0 = V_0/L$,
and the `background' potential at the cylinder location is given by
$\phi(h)=V_0(1-h/L)$.  In order to maintain a potential $V_c$, an average
zeta potential,
\be
\zeta_i = V_c - V_0\left(1-\frac{h}{L}\right),
\ee
is induced via a net transfer of charge per unit length of $\lambda_i = 2
\pi \eps \kappa a \zeta_i$, along with an equally and oppositely
charged screening cloud.

This induced screening cloud is driven by the tangential electric field (\ref{eq:steadytangentialfield})
in the standard way, giving a fixed-potential ICEO flow with slip velocity
\be
\bu_s^{{\rm FP}} = \bu_s+ 2\frac{\eps}{\eta}\frac{V_0}{L} \sin \theta\left(V_c-V_0+\frac{V_0 h}{L}\right)\thh,
\label{eq:fixedpotflow}
\ee
where $\bu_s$ is the quadrupolar
`fixed-total-charge' ICEO flow $\bu_s$ from (\ref{eq:cylinderslip}) and 
(\ref{eq:natscale}), with $E_0 = V_0/L$.  Note that both the magnitude and direction of the flow can be 
controlled by changing the position $h$ or the potential $V_c$ of the inducing conductor.
A freely-suspended cylinder would move with an electrophoretic velocity
\be
U_E =\frac{dh}{dt}=  \frac{\eps}{\eta}\frac{V_0}{L}\left(V_c-V_0+\frac{V_0
h}{L}\right) =\frac{h-h_c}{a} U_0, 
\label{eq:fixedpotvel}
\ee
away from the position $h_c = L(1-V_c/V_0)$ where its
potential is equal to the (unperturbed) background potential.  The 
velocity scale is the same as for fixed-total-charge ICEO,
although the cylinder-electrode separation $h$, rather than the
cylinder radius $a$, provides the geometric length scale.  Since
typically $h \gg a$, fixed-potential ICEO velocities are larger
than fixed-total-charge ICEO for the same field.

\begin{figure}
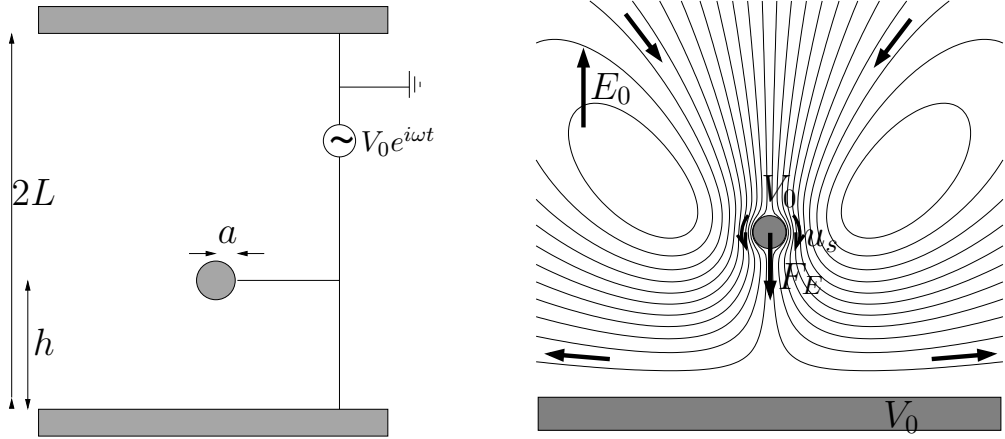

\begin{center}
\centerline{
\input{fixedpotcirc.pstex_t}
\hfill
\input{fixedpot.pstex_t}}
\caption{\label{fig:fixedpot} Fixed-potential ICEO.   a) A
cylinder of radius $a$ is held a distance $h\gg a$ from a nearby electrode
and held at the same electrostatic potential $V_0$ as the electrode.  A
second electrode is located a distance $L$ away and held at
zero potential, so that a field $E_0 \approx V_0/L$ is established.  b) Leading-order fixed-potential
ICEO flow for the system on the left.  Like fixed-charge ICEO, a non-zero steady flow can be 
driven with an AC voltage.}
\end{center}
\end{figure}
To hold the cylinder in place against $U_E$, however, a force is required.  Following \cite{jeffrey81}, the force $F_E$ 
is given approximately by
\be
F_E = \frac{4 \pi \eta U_E}{\log\left[(h+\sqrt{h^2-a^2})/a\right]-\sqrt{h^2-a^2}/a}\approx \frac{4 \pi \eta U_E}{\log\left(2h/a\right)-1}
\label{eq:fixedpotforce},
\ee
and is directed toward $h_c$.  The fixed-potential ICEO flow around a cylinder held in place at the same potential
as the nearest electrode ($V_c=V_0$, $h_c = 0$) is shown in figure \ref{fig:fixedpot}.  The leading-order flow consists of a Stokeslet of strength $F_E$ plus its images, following \cite{liron81}.  The (quadrupolar) fixed-total-charge
ICEO flow exists in addition to the flow shown, but is smaller by a factor $a/h$ and is thus not drawn.

With the ability to actively control the potential of the `inducing'
surface, fixed-potential ICEO flows afford significant additional
flexibility over their fixed-total-charge (and also colloidal) counterparts.
Note that in a sense, there is little distinction between `inducing'
conductors and blocking electrodes.  Both impose voltages, undergo time-dependent screening, and may drive
ICEO flows.  Furthermore, their sensitivity to device geometry and
nonlinear dependence on applied fields open intriguing microfluidic
possibilities for ICEO flows -- both fixed-total-charge and fixed-potential.

\subsection{ Simple microfluidic devices exploiting ICEO }
\label{sec:devices}

Owing to the rich variety of their associated phenomena, ICEO flows
have the potential to add a significant new technique to the
microfluidic toolbox.  Below, we present several ideas for
microfluidic pumps and mixers based on simple ICEO flows around
conducting cylinders.  The devices typically consist of strategically
placed metal wires and electrodes.  As such, they can be easily
fabricated and operate with no moving parts under AC applied electric
fields.  AC fields have several advantages over DC fields: (i)
electrode reactions are not required to apply AC fields, thus
eliminating the concentration and pH gradients, bubble formation and
metal ion injection that can occur with DC fields, and (ii) strong
fields can be created by applying small voltages over small distances.
For ease of analysis, we assume the cylinders to be long enough that the flow is effectively two-dimensional.

\subsubsection{ Junction pumps }

\begin{figure}
\begin{center}
\centerline{
\includegraphics[height=1.8in]{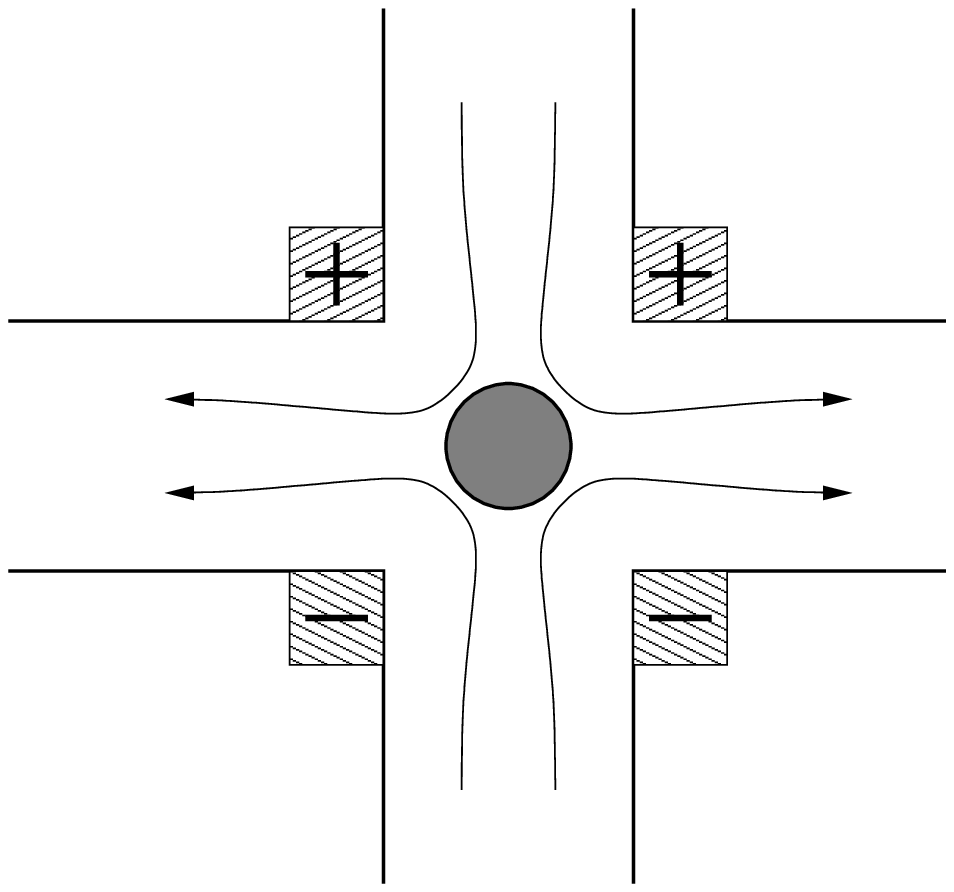}
\vspace{0.1in}
\includegraphics[height=1.8in]{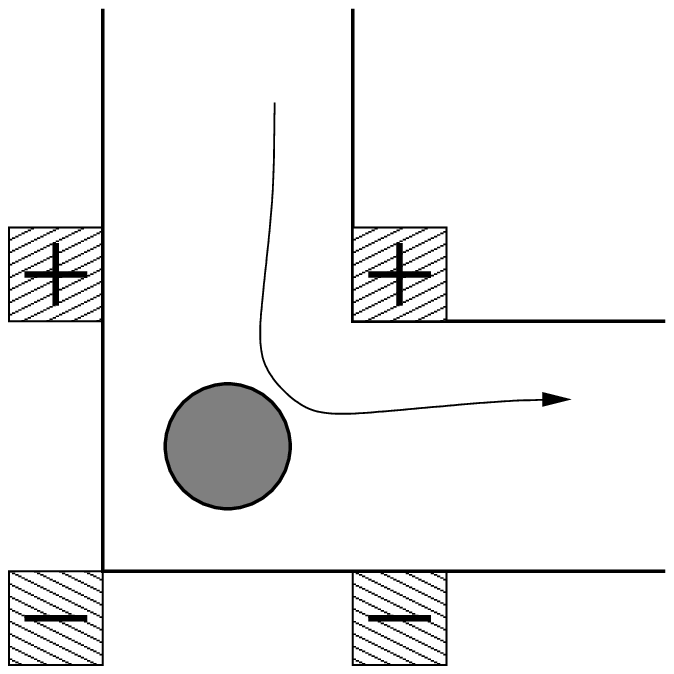}
\vspace{0.1in}
\includegraphics[height=1.8in]{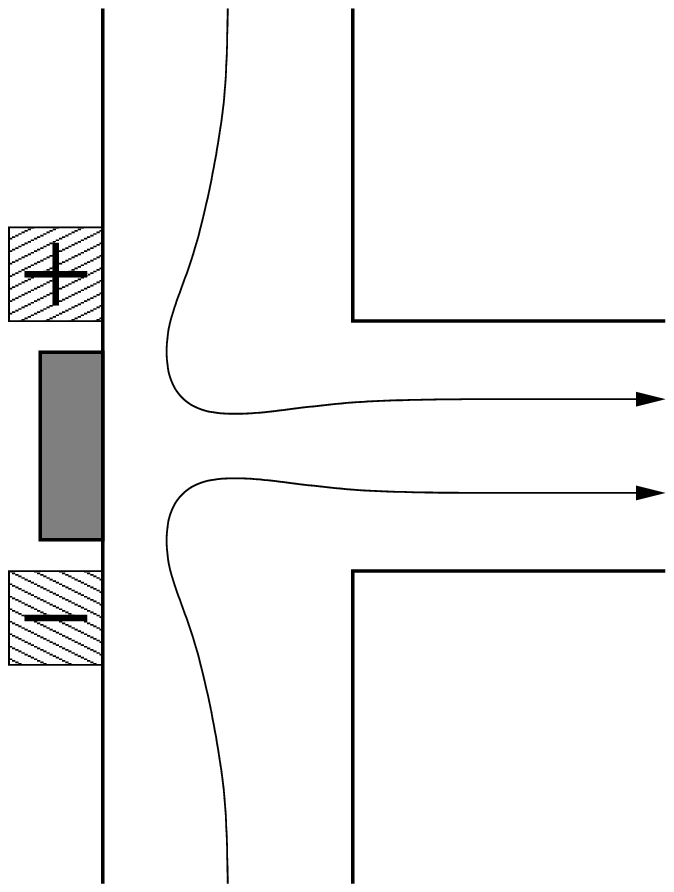}
}
\caption{\label{fig:pumpsketches} ICEO micropump designs for `cross',
`elbow', and `T' junctions. A conducting cylinder placed in a junction
of microchannels, subject to an applied AC or DC field, drives an ICEO
flow which draws fluid in along the field axis and expels it
perpendicular to the field axis.  For the four-electrode
configurations, the field axis can be switched from vertical to
horizontal by switching the polarities of two diagonally opposite
electrodes, reversing the pumping direction. }
\end{center}
\end{figure}

The simplest ICEO-based devices follow natually from the symmetry of
ICEO flows, which generally draw fluid in along the field axis and
eject it radially.  This symmetry can be exploited to drive fluid
around a corner where two, three, or four micro-channels converge at
right angles, as shown in figure \ref{fig:pumpsketches}.  These DC or
AC electro-osmotic pumps are reversible: changing the polarity of the
four electrodes (shown for the elbow and cross junctions in the
figure) so as to change the field direction by 90$^\circ$, reverses
the sense of pumping.

For variety, we give an alternate design for the T-junction pump,
which uses a conducting plate embedded in the channel wall between the
electrodes and cannot be reversed.  A reversible T junction, similar
to the cross and elbow junctions, could easily be designed. The former
design, however, can be modified to reduce the detrimental effects of
viscous drag by having the metal plate wrap around to the top and
bottom walls (not shown), in addition to the side wall. In general,
placing the `inducing conductor' driving ICEO flow on a channel wall
is advantageous because it eliminates an inactive surface that would
otherwise contribute to viscous drag.

The pressure drop generated by an ICEO junction pump can be estimated
on dimensional grounds.  The natural pressure scale for ICEO is $P
\sim \eta U_0/a \sim \eps E_0^2$, and the pressure decays with
distance like $(a/r)^{2}$.  Thus a device driven by a cylinder of
radius $a$ in a junction with channel half-width $W$ creates a pressure
head of order $\Delta P \sim \eps E_0^2 a^2/W^2$.  For the
specifications listed in Table \ref{tab:repvals}, this corresponds to
a pressure head on the order of mPa.  This rather small value suggests
that straightforward ICEO pumps are better suited for local fluid
control than for driving fluids over significant distances.



\subsubsection{ICEO micro-mixers }

\begin{figure}
\begin{center}
\includegraphics[width=2.3in]{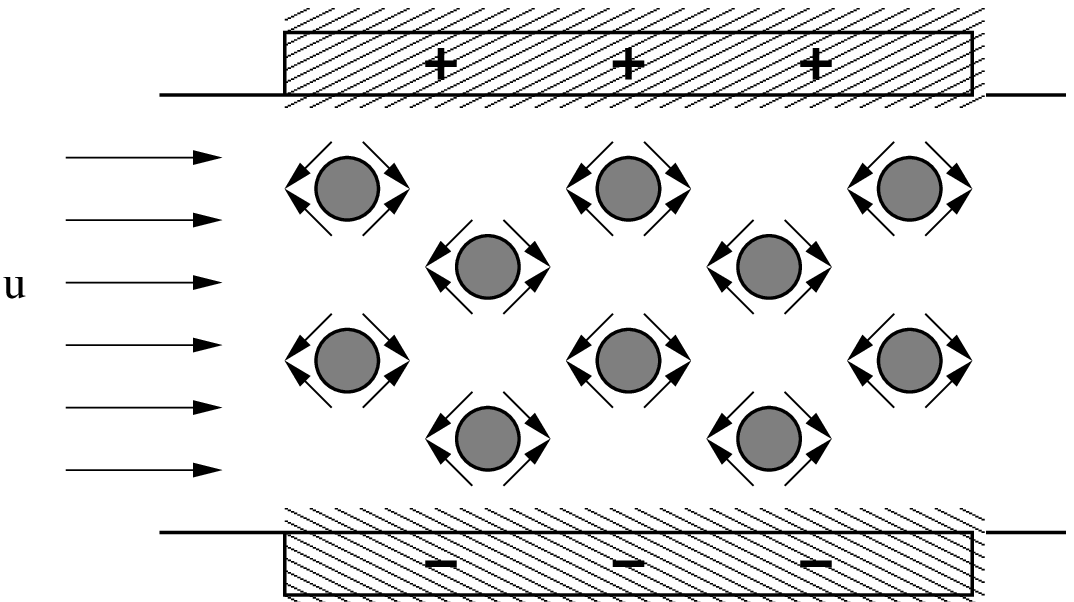}
\includegraphics[width=2.7in]{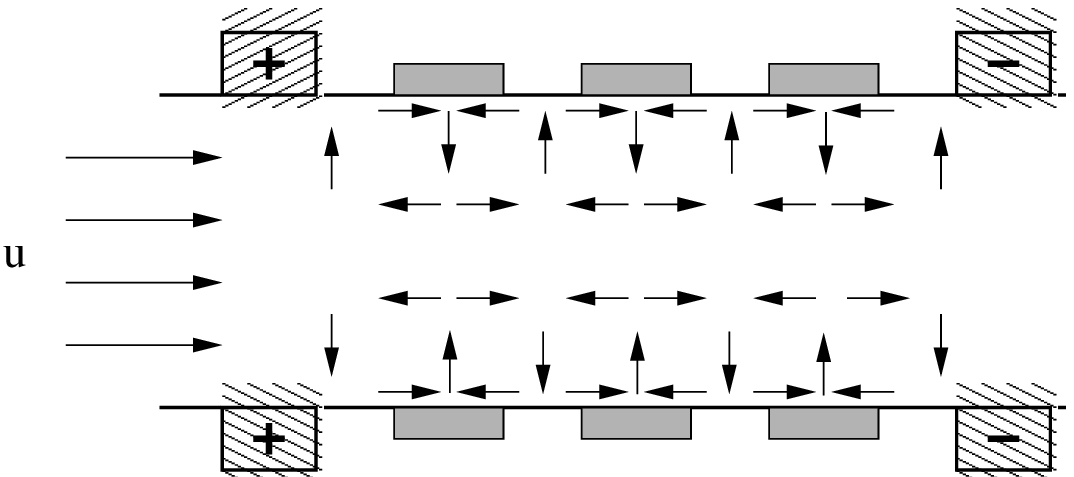}
\caption{\label{fig:mixerposts} AC electro-osmotic mixers.  Diffusive 
mixing in a background flow is enhanced by the ICEO convection rolls
produced by (a) an array of conducting posts in a transverse AC field, 
and (b) conducting objects (or coatings) embedded in channel walls 
between micro-electrodes applying fields along the flow direction. }
\end{center}
\end{figure}

As discussed above, rapid mixing in microfluidic devices is 
not trivial, since inertial effects
are negligible and mixing can only occur by diffusion. 
Chaotic advection (\cite{aref84}) provides a promising strategy for mixing
in Stokes flows, and various techniques for creating
chaotic streamlines have been introduced (\eg \cite{liu00,stroock02a}).  ICEO flows provide a simple 
method to create micro-vortices, and could therefore be used in a 
pulsed fashion to create unsteady two-dimensional flows with chaotic trajectories.

In figure \ref{fig:mixerposts}a, we present a design for an ICEO mixer
in which a background flow passes through an array of transverse
conducting posts. An AC field in the appropriate frequency range
($\tau_W^{-1} \leq \omega \leq \tau_a^{-1}$) is applied perpendicular to the
posts and to the mean flow direction, which generates an array of
persistent ICEO convection rolls.  Note that the radius of the posts
in the mixer should be smaller than shown ($a \ll W$) to validate the
simple approximations made above, where a `background' field is
applied to each post in isolation. However, larger posts as shown
could have useful consequences, as the
final field is amplified by focusing into a smaller region. We leave a
careful analysis of such issues for future work.

The same kind of convective mixing could also be produced by a
different design, illustrated in figure \ref{fig:mixerposts}b, in which
an AC (or DC) field is applied along the channel with metal stripes
embedded in the channel walls. As with the posts described above, the
metal stripes are isolated from the electrodes applying
the driving field.  This design has the advantage that it drives flow 
immediately adjacent to the wall, which reduces `dead space'.

\begin{figure}
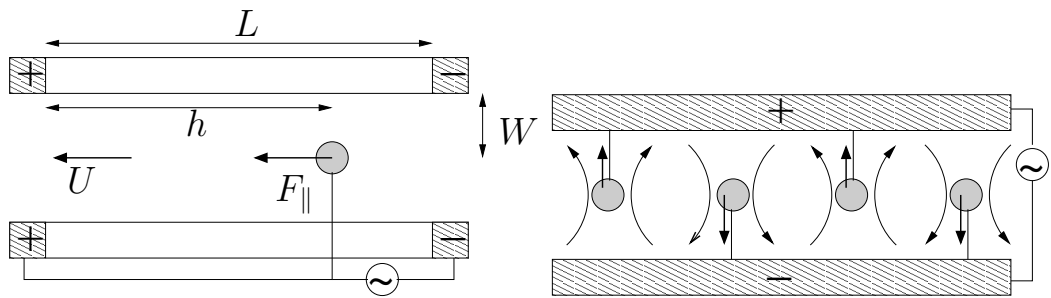

\begin{center}
\centerline{
\mbox{
\input{fixedpotpump.pstex_t}}
\ \nolinebreak
\mbox{
\input{fixedpotmixer.pstex_t}}
}
\caption{\label{fig:fixedpotpump} Fixed-potential ICEO pumps (left)
and mixers (right) can be constructed by electrically coupling the
`inducing conductor', which may be a post (as shown) or a surface
pattern (not shown), to one set of electrodes. These devices
generalize `AC electro-osmosis' in flat-surface electrodes arrays  
to other situations, with different flow patterns and
frequency responses. }
\end{center}
\end{figure}

\subsubsection{ Devices exploiting fixed-potential  ICEO}

By coupling the potential of the posts or plates in the devices above
to the electrodes, fixed-potential ICEO can be exploited to generate
net pumping past the posts.  For example, in the cross-junction pump
of figure \ref{fig:pumpsketches}, if the central post were grounded to
one of the pairs of electrodes, there would be an enhanced flow
sucking fluid in from the channel between the electrode pair (with an
AC or DC voltage).  Likewise, a fixed-potential ICEO linear pump can be created in the middle of a
channel, as shown in figure \ref{fig:fixedpotpump}a.  

To estimate the flow generated by the single-post device, we note that
the post in figure \ref{fig:fixedpotpump}a, if freely suspended, would move with velocity
$U_E \sim \eps V_0^2 h /\eta L^2$ down the channel.  The post is held in place, however,
which requires a force (per unit length) $F_\| \sim 4 \pi \eta U_E / (\log(W/a)-0.9)$
(\cite{happel83}).  
The resulting flow rate depends on the length of the channel; however,
the pressure required to stop the flow does not.  This can be estimated
using a two-dimensional analog of the calculation of \cite{brenner58} 
for the pressure drop due to a small particle in a cylindrical tube, giving a pressure
drop $\Delta P \sim 3 F_\|/4W$.  This pressure drop is larger than that for
the fixed-total-charge ICEO junction pumps by a factor of $\oo{hW^3/a^2L^2}$.
Furthermore, multiple fixed-potential ICEO pumps could be placed in series.
When the post is small ($a\ll W,L$), these pumps operate in the same frequency range $\tau_W^{-1} < \omega < \tau_c^{-1},$ as the 
junction pumps above.  

Clearly, many other designs are possible, which could provide detailed
flow control for pumping or micro-vortex generation. For example,
fixed-potential ICEO can also be used in a micromixer design, as shown
in figure \ref{fig:fixedpotpump}b, wherein rolls the size of the
channel can be established.  An interesting point is that frequency
response of the ICEO flow is sensitive to both the geometry and
the electrical couplings. We leave the design, optimization, and
application of real devices for future work, both experimental and
theoretical.

We close this section by comparing these new kinds of devices with 
previous examples of ICEO in microfluidic devices, which involve
quasi-planar micro-electrode arrays to produce `AC electro-osmosis'
(Ramos {\it et al.} 1999; Ajdari 2000).  As noted above, when the
potential of the `inducing conductor' (post, stripe, etc.) is coupled
to the external circuit, it effectively behaves like an electrode, so
fixed-potential ICEO is closely related to AC
electro-osmosis. Of course, it shares the same fundamental physical
mechanism, which we call `ICEO', as do related effects in polarizable
colloids that do not involve electrodes. Fixed-potential ICEO devices,
however, represent a significant generalization of AC electro-osmotic planar
arrays, because a nontrivial distinction arises between `electrode'
(applying the field) and `inducing conductor' (producing the primary
ICEO flow) in multi-dimensional geometries. This allows a considerable 
variety of flow patterns and frequency responses. In
contrast, existing devices using AC electro-osmosis peak at a
single frequency and produce very similar flows (Ramos {\it et al.}
1998, 1999; Brown {\it et al.}  2001; Studer {\it et al.} 2002; Mpholo
{\it et al.} 2003).

\section{Surface contamination by a dielectric coating}
\label{sec:dielectric}

The above examples have focused on an idealized situation with a clean
metal surface.  In this section, we examine the effect of a
non-conducting dielectric layer which coats the conductor, and find
that any dielectric layer which is thicker than the screening length
$\lambda_D$ significantly reduces the strength of the ICEO flow.
Furthermore, the ICEO flow around a {\sl dielectric} object, rather
than a perfectly conducting object as we have discussed so far, is
presented as a limiting case of the analysis in this section.

We start with a simple physical picture to demonstrate the basic effect of a thin dielectric layer.
Consider a conducting cylinder of radius $a$ coated with a dielectric layer of thickness $\lambda_d\ll a$ (so
that the surface looks locally planar) and
permeability $\etw$, as shown in figure \ref{fig:dielectric}.  In steady state, the potential drop from the conducting surface $\phi=0$ to the potential $\phi_\infty$
outside of the double layer occurs across in two steps:  across the dielectric (where $E=E_d$),
and across the screening cloud (where $E=E_w$), so that
\be
E_d \lambda_d + E_w \lambda_D = \phi_\infty.
\ee
\begin{figure}
\begin{center}
\centerline{
\input{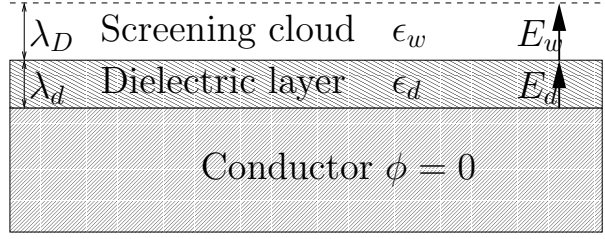}}
\caption{\label{fig:dielectric}  A dielectric layer of thickness $\lambda_d$ and permittivity $\etw$ coating
a conductor.  The potential drop between the external potential $\phi_\infty$ and 0 at the conductor occurs
in two steps:  $\Delta \phi_d=E_d \lambda_d$ across the dielectric and $\Delta \phi_w \approx E_w \lambda_D$ across
the double layer.}
\end{center}
\end{figure}
The electric fields in the double layer and in the dielectric layer are related via $\etw E_d = \eps E_w$,
so that
\be
\left(1 + \frac{\eps}{\etw} \frac{\lambda_d}{\lambda_D}\right) E_w \lambda_D = \phi_\infty.
\ee
Since $E_w \lambda_D$ is approximately the potential drop $\zeta$ across the double-layer, and since the steady-state
bulk potential is given by (\ref{eq:finalfield}) to be $\phi_\infty = 2 E_0 a \cos \theta$,
we find the induced-charge zeta potential to be
\be
\zeta = \frac{2 E_0 a \cos\theta}{1 + \eps\lambda_d/\etw \lambda_D}.
\label{eq:thindielectric}
\ee
Thus unless the layer thickness $\lambda_d$ is much less than $\etw \lambda_D/\eps$, the bulk of the potential
drop $\phi_\infty$ occurs across the dielectric layer, instead of the double-layer, resulting in a reduced electro-osmotic slip velocity.

The modification to the charging time $\tau_c$ for the coated cylinder can likewise be understood from this picture. The dielectric layer represents an additional (parallel-plate) capacitor of separation $\lambda_d$ and filled with a dielectric $\etw$, in series with the capacitive screening cloud, giving a total capacitance
\be
C_T = \frac{\eps}{\lambda_D} \left(1+\frac{\eps \lambda_d}{\etw \lambda_D}\right)^{-1},
\ee
and a modified RC time
\be
\tau_c = \frac{\lambda_D a}{D} \left(1+\frac{\eps \lambda_d}{\etw \lambda_D}\right)^{-1},
\label{eq:newrctime}
\ee 
as found by \cite{ajdari00} for a similar calculation for a compact (Stern) layer.  A discussion of
double-layer capacitance in the nonlinear regime, where the compact
layer is approximated by a thin dielectric layer is given by \cite{macdonald54}.

A full calculation of the induced zeta potential around a dielectric
cylinder of radius $a$ coated by another dielectric layer with outer
radius $b$ and thickness $(b-a)$ is straightforward although
cumbersome.  (Note that the analogous problem for a coated conducting
sphere was treated by \cite{dukhin86b} as a model of a dead biological
cell.)  The resulting induced-charge zeta potential is 
\be 
\z =
\frac{2 b E_0 (1+\Gamma_c)\cos\theta}{2+ \kappa b(1-\Gamma_c)},
\label{eq:zetacyl}
\ee
with characteristic charging time scale
\be
\tau_c =  \frac{\lambda_D b}{D }\left[1+\frac{\kappa b}{2}(1-\Gamma_c)\right]^{-1},
\ee
where $\Gamma_c$ is defined to be
\be
\Gamma_c = \frac{b^2+a^2-\eps/\etw (b^2-a^2)}{b^2+a^2+\eps/\etw (b^2-a^2)}.
\ee

It is instructive to examine limiting cases of the induced zeta potential
(\ref{eq:zetacyl}).  In the limit of a `conducting' dielectric
coating $\etw/\eps  \rightarrow \infty$, we recover the standard
result for ICEO around a metal cylinder,  as expected. In the limit of a
thin dielectric layer $b = a+\lambda_d$, where $\lambda_d \ll a$, the
induced zeta potential is given by
\be
\z (\lambda_d \ll a) \approx \frac{2 b E_0 \cos\theta}{1+\lambda_d\eps/\lambda_D \etw},
\ee
as found in (\ref{eq:thindielectric}), with a charging time
\be
\tau_c(\lambda_d \ll a) \approx \frac{\lambda_D b}{D}\left(1+\frac{\eps\lambda_d}{\etw \lambda_D}\right)^{-1},
\ee
as found in (\ref{eq:newrctime}).  Therefore, the ICEO slip velocity
around a coated cylinder is close to that of a clean conducting cylinder
($\z = 2 b E_0\cos\theta$) only when the dielectric layer is much thinner than the
screening length times the dielectric contrast, $\lambda_d \ll \lambda_D
\etw/\eps$. The zeta potential induced around a conducting cylinder with a
thicker dielectric layer,
\be
\z (\lambda_d \gg \lambda_D) \approx 2 b E_0\frac{\etw\lambda_D}{\eps\lambda_d}\cos\theta,
\ee
is smaller by a factor of $\oo{\lambda_D/\lambda_d}$, and the charging time,
\be
\tau_c(\lambda_d \gg \lambda_D) \approx\frac{\eps\lambda_D}{\etw\lambda_d}\frac{\lambda_D b}{D },
\ee
is likewise shorter by a factor of $\oo{\lambda_D/\lambda_d}$. As a
result, strong ICEO flow requires a rather clean and highly polarizable
surface, with minimal non-conducting deposits.

Note that the limit of a pure dielectric cylinder of radius $b$ is found by
taking the limit $a \rightarrow 0$. This gives a zeta potential of
\be
\z(a= 0) \rightarrow \frac{2 b E_0\cos \theta}{1+\eps b/\etw \lambda_D}\approx2\frac{\etw}{\eps} E_0\lambda_D\cos \theta,
\label{eq:dielectriccylinder}
\ee
and an ICEO slip velocity which is smaller than the conducting case
(\ref{eq:cylinderslip}) by $\oo{\lambda_D/b}$.  The charging time for a
dielectric cylinder is given by
\be
\tau_c(a= 0) \approx \frac{\eps}{\etw}\frac{\lambda_D^2}{D},
\ee
as expected from (\ref{eq:newrctime}) in the limit $\lambda_d \gg \lambda_D$.

\section{Induced-charge electro-osmosis:  systematic derivation}
\label{sec:rigorous}

In this section, we provide a systematic derivation of induced-charge electro-osmosis,
in order to complement the physical arguments above.  We derive a set of effective equations for the
time-dependent ICEO flow around an
arbitrarily-shaped conducting object, and indicate the conditions
under which the approximations made in this article are valid.
Starting with the usual `electrokinetic equations' for the
electrostatic, fluid, and ion fields (as given by, \eg,
\cite{hunter00}), we propose an asymptotic expansion that matches an
inner solution (valid within the charge cloud) with an outer region
(outside of the charge cloud), and that accounts for two separate time
scales--the time for the charge cloud to locally equilibrate and the
time scale over which the external electric field changes.
  
Although it has mainly been applied to steady-state problems, the
method of matched asymptotic expansions is well established in this
setting, where it is commonly called the `thin double layer
approximation'. 
For simplicity, like many
authors, we perform our analysis for the case of a symmetric binary
electrolyte with a single ionic diffusivity in a weak applied field,
and we compute only the leading-order uniformly valid
approximation. The same simplifying assumptions were also made by
\cite{gonzalez00} in their recent asymptotic analysis of AC
electro-osmosis.

In the phenomenological theory of diffuse charge in dilute
electrochemical systems, the electrostatic field obeys Poisson's equation,
\be
\nabla^2 \phi = -\frac{(n_+ -n_-)e}{\eps},\label{eq:govpois}
\ee
where $n_\pm$ represent the local number densities of positive and
negative ions. These obey conservation equations 
\be
\pd{n_\pm}{t} + \nabla \cdot (n_\pm \bv_\pm)=0,\label{eq:govionconservation}
\ee
where $\bv_\pm$ represent the velocities of the two ion species,
\be
\bv_\pm = \mp b  e \nabla \phi - k_B T b\nabla \log n_\pm + \bu, 
\label{eq:govionvelocity}
\ee
where $b$ is the mobility of the ions in the solvent and $\bu$ is the
local fluid velocity.  Terms in (\ref{eq:govionvelocity})
represent ion motion due to (a) electrostatic forcing, (b) diffusion
down density gradients, and (c) advection with the local fluid
velocity.  The fluid flow obeys the Stokes
equations, with body force given by 
the product of the charge density with the electric field,
\be
\eta \nabla^2 \bu - \nabla p = e(n_+ - n_-) \nabla \phi, \label{eq:govnavstokes}
\ee
along with incompressibility.  Strictly speaking, this form does not explicitly include 
osmotic pressure gradients, as detailed below.  However, osmotic forces can be 
absorbed into a modified pressure field $p$, and the resulting flow is unaffected.

For boundary conditions on the surface $\Gamma$ of the conductor, we
require the fluid flow to obey the no-slip condition, the electric
potential to be an equipotential (with equilibrium zeta potential
$\zeta_0$), and the ions to obey a no-flux condition:
\bea
\bu(\Gamma) &=& {\bf 0}\label{eq:noslip}\\
\phi(\Gamma) &=& \zeta_0 \label{eq:equipotential}\\
 \bnh \cdot \bv_\pm(\Gamma)&=& 0, \label{eq:govionbc}
\eea
where $\bnh$ represents the (outer) normal to the surface $\Gamma$.
Far from the object, we require the fluid flow to decay to zero, the
electric field to approach the externally-applied electric field, and
the ion densities $n_\pm$ to approach their constant (bulk) value
$n_0$. 

In order to simplify these equations, we insert
(\ref{eq:govionvelocity}) in (\ref{eq:govionconservation}), and take
the sum and difference of the resulting equations for the two ion
species to obtain
\bea
\ddendif + D\kappa^2 \dendif - D\nabla^2 \dendif - e b\nabla \cdot \left[\densum\nabla \phi\right]+ \bu \cdot \nabla \dendif &=&0\label{eq:fullionequation}\\
\ddensum - D\nabla^2 \densum - e b \nabla \cdot \left[\dendif \nabla \phi\right] + \bu \cdot \nabla \densum &=&0\label{eq:fullionequationalpha},
\eea
where we have used (\ref{eq:govpois}) and where we have defined
\bea
\densum &=& n_++n_- -2n_0 \\
\dendif &=& n_+-n_-.
\eea
The first variable, $\densum$, represents the
excess total concentration of ions,
while the second, $\dendif$, is related to the
charge density via $\dendif=\rho/e$.
The first three terms in (\ref{eq:fullionequation}) represent a
(possible) transient, electrostatic transport, and diffusive
transport, respectively, and will be seen to give the dominant
balance in the double layer. The next term represents the divergence of a flux of excess
ionic concentration $\densum$ (but not excess charge) due to an
electric field.
The final term represents ion advection with the fluid flow $\bu$.
The analogous (\ref{eq:fullionequationalpha}) for the charge-neutral
$\densum$ lacks the electrostatic transport term.
Both $\densum$ and $\dendif$ decay away from the solid surface and obey the no-flux boundary conditions (\ref{eq:govionbc}),
\bea
\nh\cdot\nabla \dendif|_\Gamma  &=& -\frac{e}{k_B T}(2 n_0 + \densum) \nh\cdot\nabla \phi |_\Gamma  \label{eq:deltabc}\\
\nh\cdot\nabla \densum|_\Gamma  &=& -\frac{e}{k_B T}\dendif \nh\cdot\nabla  \phi |_\Gamma   \label{eq:alphabc},
\eea
at the surface $\Gamma$.

In what follows, we obtain an approximate solution to the governing equations
(\ref{eq:govpois}), (\ref{eq:fullionequation}), and
(\ref{eq:fullionequationalpha}), at the leading order in a matched
asymptotic expansion.  We first analyze the solution in the `inner'
region within a distance of order $\lambda_D$ of the surface.
Non-dimensionalization yields a set of approximate equations for the
inner region and a set of matching boundary conditions which depend on
the `outer' solution, valid in the quasi-neutral bulk region (farther
than $\lambda_D$ from the surface). Similarly, approximate equations and
effective boundary conditions for the outer region are found. By
solving the inner problem and matching to the outer problem, a set of
effective equations are derived for time-dependent, bulk ICEO flows.

\subsection{Non-dimensionalization and `inner' region}

We begin by examining the `inner region', adjacent to the conducting surface.  Denoting non-dimensional variables in the inner region with tildes, we scale variables as follows,
\be
\br = \lambda_D \nr, \ \  t =  (\kappa^2 D)^{-1} \nt, \ \ \phi = \Phi_0 \tilde{\phi}, \ \ \bu = U_0 \bnu,\ \dendif = 2 n_0 \Psi \ndendif, \ \densum = 2 n_0 \Psi^2 \ndensum,\ \ \bE = E_0\tilde{\bE}
\ee
where $\Phi_0$ and $U_0$ are potential and velocity scales (left unspecified for now), and where
we have introduced the dimensionless surface potential,
\be
\Psi = \frac{e \Phi_0}{k_BT}.\label{eq:defpsi}
\ee
Note that $\ndensum$ scales with $\Psi^2$ to satisfy the dominant balance in (\ref{eq:fullionequationalpha}).  Finally, we have scaled time with the Debye time, $\tau_D = (\kappa^2 D)^{-1} = \lambda_D^2/D$, although the 
analysis will dictate another time scale, as expected from the physical arguments above.  

The dimensionless ion conservation equations (\ref{eq:fullionequation}-\ref{eq:fullionequationalpha}) become
\bea
\pd{\ndendif}{\tilde{t}}+\ndendif - \nnabla^2 \ndendif - \Psi^2 \nnabla \cdot \left[\ndensum\nnabla \nphi\right] - {\rm Pe}\, \bnu \cdot \nnabla \ndendif &=&0\label{eq:deltaeqn}\\
\pd{\ndensum}{\tilde{t}}+ \nnabla^2 \ndensum + \nnabla \cdot \left[\ndendif \nnabla \nphi \right] - {\rm Pe}\, \bnu \cdot \nnabla \ndensum &=&0\label{eq:alphaeqn},
\eea
where we have introduced the P\'eclet number,
\be
\Pe = \frac{U_0}{\kappa D}.\label{eq:defp}
\ee

The boundary conditions (\ref{eq:deltabc}-\ref{eq:alphabc}) in non-dimensional form are given by
\bea
\nh\cdot\nnabla \ndendif|_\Gamma &=& -(1+\Psi^2\ndensum)\nh\cdot\nnabla\nphi |_\Gamma \label{eq:boundarydelta}\\
\nh\cdot\nnabla \ndensum|_\Gamma &=& -\ndendif \nh\cdot\nnabla\nphi |_\Gamma\label{eq:boundaryalpha}.
\eea

In the analysis that follows, we concentrate on the simplest limiting case.
We assume the screening length to be much smaller than any length
$L_0$ associated with the surface geometry, parametrized through
\be 
\epsilon = (\kappa L_0)^{-1} = \lambda_D/L_0 \ll 1,
\label{eq:defep}
\ee 
so that the screening cloud `looks' locally planar.  As mentioned
above, the singular limit of thin double layers, $\epsilon \ll 1$, is the
usual basis for the matched asymptotic expansion.  The regular limit
of small P\'eclet number, $\Pe \ll 1$, which holds in almost any
situation, is easily taken by setting $\Pe=0$.  Finally, we consider the
regular limit of small (dimensionless) surface potential, $\Psi \ll
1$, which is the same limit that allows the Poisson-Boltzmann equation
to be linearized.  With these approximations, the system is
significantly simplified: First, $\ndendif$ is coupled to $\ndensum$
only through terms of order $\oo{\Psi^2}$ in (\ref{eq:deltaeqn})
and boundary condition (\ref{eq:boundarydelta}).  Second,
$\ndensum$ is smaller than $\ndendif$ by a factor $\Psi$, and is thus
neglected: $\ndendif = 0 + \oo{\Psi^2}$.

In this limit, we obtain the linear Debye equation and
boundary condition for $\ndendif$ alone: 
\bea
\pd{\ndendif}{\tilde{t}}+\ndendif = \epsilon^2 \nnabla^2 \ndendif
\label{eq:simpleion}
\\
\nh\cdot\nnabla \ndendif|_\Gamma = -\nh\cdot\nnabla\nphi |_\Gamma.
\label{eq:simpleionbc}
\eea 
The potential is then recovered from 
(\ref{eq:govpois}) in the form,
\be
\nnabla^2 \nphi = -\ndendif,
\label{eq:ndpois}
\ee
with the far-field boundary condition,
\be
\nnabla \nphi \rightarrow - \frac{E_0 L}{\Phi_0}\tilde{\bE} ,\label{eq:farfieldbc}
\ee
and the (equipotential) surface boundary condition 
\be
\nphi(\Gamma) = \Phi_0 \nzeta_0\label{eq:ndpoisbc}.
\ee
where $\nzeta_0$ is the dimensionless equilibrium zeta potential. 

Since $\ep\ll 1$, we introduce a locally 
Cartesian coordinate system $\{\nn,\nl\}$, where $\nn$ is locally normal to the
surface, and $\nl$ is locally tangent to the surface. 
The governing equations (\ref{eq:simpleion}) and
(\ref{eq:ndpois}) are both linear in $\ndendif$ and $\nphi$, which allows the
electrostatic and ion fields to  be expressed as a simple
superposition of the equilibrium fields,  
\be
\ndendif_{\rm eq} = -\nphi_{\rm eq} = \nzeta_0 e^{-\nn},
\ee
and the time-dependent induced fields  $\ndendif_{\rm i}$ and $\nphi_{\rm i}$, via
\bea
\ndendif = \ndendif_{\rm eq}+\ndendif_{\rm i}\\
\nphi = \nphi_{\rm eq}+\nphi_{\rm i}.
\eea
The induced electrostatic and ion fields must then obey
\bea
\nnabla^2 \nphi_i = -\ndendif_i \label{eq:solvees}\\
\pd{\ndendif_i}{\tilde{t}}+\ndendif_i - \nnabla^2 \ndendif_i =0,\label{eq:solveion}
\eea
subject to boundary conditions
\bea
\nphi_i(\Gamma) &=& 0 \label{eq:solveesbc}\\
\nh\cdot\nnabla \ndendif_i|_\Gamma &=& -\nh\cdot\nnabla\nphi_i |_\Gamma,\label{eq:solveionbc}
\eea
on the surface.

Another consequence of the linearity of (\ref{eq:simpleion}) and
(\ref{eq:ndpois}), and therefore of the decoupling of the equilibrium
and induced fields, is that different characteristic scales can be
taken for the two sets of fields.  We scale the potential in the equilibrium
problem with the equilibrium zeta potential $\zeta_0$, and
the potential in the induced problem with $E_0 L_0$, the
potential drop across the length scale of the object.  In order that
the total surface potential be small $(\Psi \ll 1)$, however, we require
that the total zeta potential be small,
\be
\frac{(|\zeta_0| + |E_0 L_0|)e}{k_B T} \ll 1.
\ee
At the end of this section, we will briefly discuss  the rich variety of
nonlinear effects which generally occur, in addition to ICEO, when
this condition is violated.

We expect the charge cloud $\ndendif$ and electric potential $\nphi$
to vary quickly with $\nn$, but slowly with $\nl$ (along the surface).
Furthermore, we expect the charge cloud to exhibit two time scales: a
fast (transient) time scale, over which $\ndendif$ reaches a
quasi-steady dominant balance, and a slow time scale
$\tilde{\tau_c}$, over which the quasi-steady solution changes.  The
physical arguments leading to (\ref{eq:chargingtime}) for the
charging time suggest that this slow time scale is given by
$\tilde{\tau}_c = 1/\ep$, which is not obvious {\sl a priori}, but
which will be confirmed by the successful asymptotic matching.

In order to focus on the long-time dynamics of the induced charge
cloud, and guided by the above expectations, we attempt a quasi-steady
solution to (\ref{eq:solvees}) -- (\ref{eq:solveionbc}) of the form
\bea
\ndendif_i = \ndendif_i(\nn,\ep \nl,\ep \nt) \label{eq:matchedexpa}\\
\nphi_i = \nphi_i(\nn,\ep \nl, \ep \nt) \label{eq:matchedexpb}.
\eea
It can be verified that
\bea
\ndendif_i &=& A(\ep \nl,\ep \nt) e^{-\nn} -\ep \frac{\dot{A}(\ep \nl,\ep \nt) }{2}\nn e^{-\nn}\label{eq:rhosol}\\
\nphi_i &=& -A(\ep \nl,\ep \nt) e^{-\nn} + \ep \frac{\dot{A}(\ep \nl,\ep \nt) }{2}\left[2 e^{-\nn}+\nn e^{-\nn}\right] + B(\ep \nl,\ep \nt) + C(\ep \nl,\ep \nt) \nn,
\eea
solve the governing equations to $\oo{\ep^2}$.  

The equipotential boundary condition (\ref{eq:solveesbc}) is satisfied when
\be
A(\ep \nl,\ep \nt) = B(\ep \nl,\ep \nt) + \ep \dot{A}(\ep \nl,\ep \nt) \label{eq:equipcond},
\ee
so that the induced double-layer charge density is proportional (to leading order in $\ep$) to 
the potential just outside the double layer.  The normal ion flux condition (\ref{eq:solveionbc}),
\be
\ep \dot{A}(\ep \nl,\ep \nt) = C(\ep \nl,\ep \nt), \label{eq:normalfield}
\ee
relates the evolution of the double-layer charge density to the electric field normal to the double-layer.
Matching the inner region to the outer region provides the final relations.  In 
the limit $\nn \rightarrow \infty$, the electrostatic and ion fields approach their limiting behavior
\bea
\ndendif &\rightarrow& 0 \label{eq:matchone}\\
\nphi &\rightarrow& B(\ep \nl,\ep \nt) + C(\ep \nl,\ep \nt) \nn \label{eq:matchtwo}.
\eea

\subsection{Outer solution}
We now turn to the region outside of the double-layer, and use overbars to
denote `outer' non-dimensional variables (scaled with length scale
$L_0$ and potential scale $E_0 L_0$).  According to (\ref{eq:rhosol})
and (\ref{eq:matchone}), any non-zero charge density $\hdendif$ that
exists within the inner region decays exponentially away from the
surface $\Gamma$.  A homogeneous solution ($\hdendif_o = 0$) thus
satisfies (\ref{eq:simpleion}) and the decaying boundary condition
at infinity.  With $\hdendif_o = 0$, (\ref{eq:ndpois}) for the
electrostatic field $\hphi_o$ in the outer region reduces to Laplace's
equation,
\be
\hnabla^2 \hphi_o = 0,\label{eq:outerlaplace}
\ee
with far-field boundary condition (\ref{eq:farfieldbc}) given by 
\be
\hnabla \hphi_o \rightarrow -\hat{\bE}.  
\ee
To determine $\hphi_o$ uniquely, one further boundary condition on the surface $\Gamma$ is required,
which is obtained by matching to the inner solution.  The limiting value of the `outer' field $\hphi_o$,
\be
\hphi_o(\hr \rightarrow \Gamma) \rightarrow \hphi_0(\Gamma) + \hat{E}_\perp(\Gamma) \hn,
\ee 
must match the inner solution (\ref{eq:matchtwo}), which gives two relations
\bea
B(\ep \nl,\ep \nt) =  \nphi_0(\nl), \label{eq:matchthree}\\
C(\ep \nl,\ep \nt) = \ep \tilde{E}_\perp(\nl) \label{eq:matchfour}.
\eea
Using (\ref{eq:matchfour}) in (\ref{eq:normalfield}), the fact that $C$ is $\oo{\ep}$ verifies that the assumed time scale $\tau_c = \tau_D / \ep$ for induced double layer evolution is indeed correct, as expected from physical arguments (\ref{eq:chargingtime}). 

\subsection{Effective equations for ICEO in weak applied fields}

Using the above results, we present a set of effective equations for
time-dependent ICEO that allows the study of the large-scale flows,
without requiring the detained inner solution.  What emerges is a
first-order ODE for the dimensionless total surface charge density in the
diffuse double layer, $\tilde{q}=A$, so an `initial' value
for $\tilde{q}$ must be specified.  From (\ref{eq:equipcond}) and
(\ref{eq:matchthree}), the `outer' potential on the surface $\Gamma$ is
given by
\be
\hphi_o(\Gamma, \ep \htt) = \tilde{q}(\Gamma, \ep \htt),
\ee
which, along with the far-field boundary condition
(\ref{eq:farfieldbc}), uniquely specifies the solution to
Laplace's equation (\ref{eq:outerlaplace}).  From this solution,
the normal field $\hat{E}_\perp(\Gamma,\ep \htt)$ is found, which
(using (\ref{eq:matchfour}) and (\ref{eq:normalfield})) results in a
time-dependent boundary condition,
\be
\frac{\partial\tilde{q}}{\partial\htt} =  \ep \hat{E}_\perp(\Gamma,\ep \htt),
\ee
which is more naturally expressed as
\be
\frac{\partial\tilde{q}}{\partial \hat{t}} =
\hat{E}_\perp(\Gamma,\hat{t}),
\label{eq:genlinbc}
\ee
in terms of the dimensionless time variable, $\hat{t} = \ep \htt =
t/\tau_c$. 

We have thus matched the bulk field outside the charge cloud with the
`inner' behavior of the charge cloud in a self-consistent manner.  The
`inner' solutions for $\dendif$ and $\zeta$ equilibrate quickly in response
to the (slow) charging, and affect the boundary conditions which determine
the `outer' solution.  Matching (\ref{eq:matchfour}) and (\ref{eq:normalfield})
results in a relation between the charge cloud and normal ionic flux,
confirming the validity of (\ref{eq:dzetadt}) and (\ref{eq:jeperp}),
which were previously argued in an
intuitive, physical manner.  This analysis has demonstrated that errors to
this approach are of order $\oo{\Psi^2}$,  $\oo{P}$ and
$\oo{\ep^2}$. In this limit, the system evolves with a single characteristic time
scale, $\tau_c = \lambda_D L_0/D$, set by asymptotic matching, which
physically corresponds to an RC coupling, as explained above.

\subsection{ Fluid dynamics }
\label{sec:fluidflow}

\subsubsection{Fluid body forces: electrostatic and osmotic}

Thus far, we have derived the effective equations for the dynamics of the 
induced double-layer.  To conclude, we examine the ICEO slip velocity that 
results from the interaction of the applied field and the induced diffuse layer.
We demonstrate that, in the limits of thin double layer and small surface potential,
the Smoluchowski formula (\ref{eq:smoluchowski}) for fluid slip holds.

Following \cite{levich62} (p. 484), we consider a conducting surface $\Gamma$ immersed in a fluid, with an applied `background' electrostatic potential $\phi_o$.  In addition, a thin double-layer (with potential $\xi$) exists and obeys
\be
\nabla^2 \xi = -\frac{(n_+ - n_-)e}{\eps} = -\frac{n_0 e}{\eps}\left(e^{-e \xi/k_B T} - e^{e \xi/k_B T}\right),
\ee
so that the total potential $\phi = \phi_o + \xi$.  Note that this assumes that the double layer is in quasi-equilibrium.

Fluid stresses have two sources:  electric and osmotic.  The electric stress in the fluid is given by the Maxwell Stress tensor,
\be
T_{ij} = \eps \left(E_i E_j - \frac{1}{2} \bE\cdot\bE \delta_{ij}\right),
\ee
from which straightforward manipulations yield the electrical body force on the fluid to be
\be
\bF_E \equiv \nabla \cdot {\bf T} = \eps \nabla \phi \nabla^2 \phi \equiv \eps \nabla (\phi_o + \xi) \nabla^2 \xi.
\label{eq:electricforce}
\ee
Osmotic stresses come from gradients in ion concentration, and exert a fluid body force 
\be
\bF_O = -k_B T \nabla \left(n_+ + n_-\right) = e \nabla \zeta \left(n_+ - n_-\right) =  - \eps \nabla \xi \nabla^2 \xi.
\label{eq:osmoticforce}
\ee
Thus the total body force on the fluid, is given by the sum of $\bF_E$ and $\bF_O$,
\be
\bF = \eps \nabla \phi_o \nabla^2 \xi = \rho {\bf E}_B.
\label{eq:stokesforcing}
\ee
Therefore, when both electric and osmotic stresses are included, the body force on the double layer above a conductor is given by the product of the local charge density $\rho$ and the `background' electric field $\bE_B = -\nabla \phi_o$ (which, importantly, does not vary across the double layer).  Note also that the same fluid flow would result if the double-layer forcing $\rho ({\bf E}_B + {\bf E}_\xi)$ were to be used, since the osmotic component  (\ref{eq:osmoticforce}) is irrotational and can be absorbed in a modified fluid pressure.

\subsubsection{ICEO slip velocity}

Finally, we examine the ICEO slip velocity that results when the electric
field $-\nabla \phi_o$ drives the ions in the induced charge screening
cloud $\rho$.  

We look first at the flow in the `inner' region of size $\lambda_D$.  
For the ICEO slip velocity to reach steady state, vorticity must diffuse across the double layer, which requires a very small time $\tau_\omega(\lambda_D) = \lambda_D^2/\nu \approx 10^{-10}$ s.  Because $\tau_\omega$ is so much faster than the charging time and Debye time, we consider the ICEO slip velocity to follow changes in $\zeta$ or $\phi$ instantaneously.  As noted above, the unsteady term may play a role in cutting off the oscillating component of the bulk flow.  However, our main concern is with the steady, time-averaged component, which simply obeys the steady, forced Stokes equation.

Non-dimensionalizing as above, we re-express the forced Stokes equations (\ref{eq:govnavstokes}), with forcing given by (\ref{eq:stokesforcing}), using a stream function $\psi$ defined so that $u_l = \partial_n \psi$ and $u_n = -\partial_l \psi$.  The stream function obeys
\be
\nnabla^4 \npsi = \left(\pd{\ndendif}{\nn}\pd{\nphi}{\nl}-\pd{\ndendif}{\nl}\pd{\nphi}{\nn}\right),
\label{eq:innerfluid}
\ee
where the stream function has been scaled by $\psi = (\eps E_0 \zeta_0/\eta \kappa) \npsi.$
We perform a local analysis around the point $\nl=0$ of the ICEO flow driven by an applied tangential electric field $\{\nE_\|,\nE_\perp\}$.  Using representations of $\ndendif$ and $\nphi$ around $\nl = 0$,
\bea
\ndendif &=& \ndendif(\ep \nl) e^{-\nn}\\
\nphi &= & -\nE_\| \nl (\ep \nl,\ep \nn) -\nE_\perp\nn (\ep \nl,\ep \nn),
\eea
it is straightforward to solve (\ref{eq:innerfluid}) to $\oo{\ep}$.  The tangential and normal flows are
then given by
\bea
u_l&=& -\frac{\eps}{\mu} \left[\left(\zeta(l) E_\|(l) + \ep \pd{\zeta}{\nl} E_\perp\right) \left(1-e^{-\kappa n}\right) + \ep \pd{E_\|}{\nn} \left(3 - (3 +\kappa n) e^{-\kappa n} \right)\right]\\
u_n &=& -\ep \frac{\eps}{\mu} \pd{}{\nl}\left(\zeta E_\|\right) \left(1-\kappa n - e^{-\kappa n}\right),
\eea
where $\zeta(l) E_\|$ contains terms of $\oo{\ep}$.  To leading order, then, the slip flows obey
\be
u_l \rightarrow -\frac{\eps}{\mu} \zeta(l) E_\|(l)+\oo{\ep} \ \ \mbox{and} \ \ 
u_n = \oo{\ep}.
\ee
The tangential flow does indeed asymptote to (\ref{eq:smoluchowski}), with local zeta potentials and background field, and the normal flow velocity is smaller by a factor of order $\oo{\ep}$.

Thus an ICEO slip velocity is very rapidly established in
response to an induced zeta potential and `outer' tangential field. 
Furthermore, despite double-layer and tangential field gradients, the 
classical Helmholtz-Smoluchowski formula, (\ref{eq:smoluchowski}),
correctly gives the electro-osmotic slip velocity.  This may
seem surprising, given that the tangential field {\sl vanishes} at the 
conducting surface.

The final step involves finding the bulk ICEO flow must be found by solving the unsteady Stokes equations,
with no forcing, but with a specified ICEO slip velocity on the boundary
$\Gamma$, given by solving the effective electrokinetic transport problem
above.

\subsection{ Other nonlinear phenomena at large voltages }
\label{sec:nonlinear}

Although we have performed our analysis in the linearized limit of small
potentials, it can be generalized to the `weakly nonlinear' limit
of thin double layers, where (\ref{eq:valid}) holds and $\Du
\ll 1$. In that case, the bulk concentration remains uniform at
leading order, and the Helmholtz-Smoluchowski slip formula remains
valid. The main difference involves the time-dependence of double-layer
relaxation, which is slowed down by nonlinear screening once the
thermal voltage is exceeded.  It can be shown that the linear time-dependent
boundary condition, (\ref{eq:dzetadt}) or (\ref{eq:genlinbc}),
must be modified to take into account the nonlinear differential
capacitance, as in (\ref{eq:nonlin_charge}).  
Faradaic surface reactions and the capacitance of the compact Stern layer may also be included
in such an approach, as in the recent work of \cite{bonnefont01}.

The condition (\ref{eq:valid}) for the breakdown of Smoluchowski's
theory of electrophoresis (with $\zeta = \zeta_0 + E_0a$) coincides
with the condition $\tau_c(\zeta) \ll \tau_D$, where $\tau_c$ takes
into account the nonlinear differential capacitance (\ref{eq:tcnonlin}). When
this condition is violated (the `strongly nonlinear' regime),
double-layer charging is slowed down so much by nonlinearity that it
continues to occur at the time scale of bulk diffusion. At such large
voltages, the initial charging process draws so much neutral
concentration into the double layer that it creates a transient
diffusion layer which must relax into the bulk, while coupled to the
ongoing double-layer charging process. 
\cite{bazant04b} explore such processes. 

In the strongly nonlinear regime, the Dukhin number is typically not
negligible, and bulk concentration gradients (and their associated
electrokinetic effects) are produced by surface conduction (see \cite{dukhin93} and
\cite{lyklema91}).  Tangential concentration gradients
modify the usual electro-osmotic slip by changing the bulk electric
field (concentration polarization) and by producing
diffusio-osmotic slip.  Therefore, both the steady state and the 
relaxation processes for ICEO flows are affected.

Finally, large voltages can also lead to the breakdown of ideal
polarizability via spontaneous Faradaic reactions on different sides
of the object, as noted by \cite{murtsovkin91b}.  \cite{gamayunov92}
observed that the induced-charge electro-osmotic flow around metal
colloidal spheres reverses direction for large colloids, and argued
that Faradaic reactions were responsible.  Furthermore,
\cite{barany98} measured large conducting colloids to have `superfast'
(second-kind) electrophoretic velocities.  They argued that
sufficiently strong fields cause Faradaic currents at the two sides of
the colloid (as though they were electrodes), resulting in the
development of a bulk `space charge' and, correspondingly, second-kind
electrophoresis.

A complete description of time-dependent ICEO at large voltages, which
is beyond the scope of this article, would require considering all of
these effects at once. The approximation of thin double layers, which has been applied
mostly to steady-state problems involving non-polarizable objects, is
a good starting point.  However, the presence of multiple length and time scales
complicates mathematical analysis, especially in any attempt to go beyond the leading-order
approximation. An important goal, therefore, would be to extend the
method of matched asymptotic expansions to derive effective equations
and boundary conditions for strongly nonlinear ICEO and to carefully
analyze asymptotic corrections.

\section{Summary and Discussion}

In this article, we have described the general phenomenon of
induced-charge electro-osmosis (ICEO), which includes a wide variety
of techniques (both old and new) for driving steady micro-flows
around conducting or dielectric surfaces using AC or DC electric
fields.  We have given a physical picture of the basic mechanism for
ICEO, involving the inhomogeneous surface charge induced in the
conductor in order to maintain an equipotential surface in the
presence of an applied field.  In response, the electric field normal
to the surface/charge cloud drives ions into an inhomogeneous
(dipolar) charge cloud, which are in turn driven by the tangential
electric field.  This results in ICEO slip velocities of magnitude
$U_0 \sim \eps E_0^2 a/\eta$.  A charging time scale $\tau_c \sim
\lambda_D a/D$ is required for these induced charge clouds to form.  Due
to the dependence on the square of the applied field $E_0$, a nonzero
time-averaged ICEO flow can be driven using AC fields of sufficiently
low frequencies $(\omega \ll 1/\tau_c)$.

We have performed explicit calculations for the steady and unsteady ICEO
slip velocities (suddenly-applied and sinusoidal fields) around symmetric
conducting cylinders.  The ICEO flow for conducting cylinders is quadrupolar and decays with
distance like $r^{-1}$.  
We have also performed a systematic,
matched-asymptotic analysis of the equations for the ion transport,
electrostatics and fluid flow to confirm the validity of the physically
intuitive approach. The analysis produces a set of effective equations
which `integrate out' the dynamics of the thin screening cloud, allowing
the bulk ICEO flow to be calculated with macro-scale calculations alone.

We have also considered polarizable dielectric  surfaces for two reasons.
First, we have shown that a dielectric layer of thickness $\lambda_d$ can
reduce the strength of the ICEO slip velocity by a factor of order
$\lambda_D/a$ when the dielectric layer is sufficiently thick $(\lambda_d
\gg \lambda_D)$. This underscores the necessity of using clean and/or
treated surfaces to ensure a clean conductor/water interface. Second, 
an ICEO flow is set up even around a purely dielectric colloidal
particle of permittivity $\etw$, but with reduced ICEO slip velocity, $U_0
\sim \etw E_0^2 \lambda_D/\eta$.

In this article, we have concentrated upon ICEO flows in systems
of high symmetry: circular cylinders and spheres in spatially uniform
applied fields, for which simple exact solutions are possible.  In a
companion article, we will explore the implications of broken spatial
symmetries -- both via asymmetric surface properties and gradients in
the applied electric field -- along with more potential applications
to microfluidic devices. For a brief summary of our results, the
reader is referred to \cite{bazant04a}.

In conclusion, ICEO is a rather general and potentially useful
phenomenon, capable of producing large fluid `slip' velocities around
polarizable surfaces, under AC or DC fields.  Many variants exist on
the basic situations presented in this article.  For example, one can
apply spatially inhomogeneous electric fields, vary the geometry or
electrical properties of the polarizable surface, apply
fixed-potential (or actively-controlled potential) ICEO flows, and so
on. The directions seem promising to pursue experimentally in real
microfluidic devices.

In the presence of bulk concentration gradients produced by surface
conduction, Faradaic processes, or transient double-layer
adsorption, more general electrokinetic phenomena may also occur at
polarizable surfaces.  These effects have been described to varying
degrees in the Russian literature on `non-equilibrium electrosurface
phenomena' in colloidal systems, especially over the past few
decades. It is our hope that this mature subject, which includes ICEO
as a limiting case, will gain renewed attention from the microfluidics
community in the coming years.

\section*{ Acknowledgments }
The authors would like to thank \'Ecole Sup\'erieure de Physique et
Chimie Industrielles (Laboratoire de Physico-chimie Th\'eorique) for
hospitality and partial support, and the referees for
extensive comments and Russian references.  This research was supported in
part by the U.S. Army through the Institute for Soldier
Nanotechnologies, under Contract DAAD-19-02-0002 with the U.S. Army
Research Office (MZB), and by the NSF Mathematical Sciences
Postdoctoral Fellowship and Lee A. Dubridge Prize Postdoctoral
Fellowship (TMS).

\appendix




\begin{thebibliography}{69}
\expandafter\ifx\csname natexlab\endcsname\relax\def\natexlab#1{#1}\fi

\bibitem[Ajdari (1995)]{ajdari95}
{\sc Ajdari, A.} 1995 Elctroosmosis on inhomogeneously charged surfaces. {\em
  Phys. Rev. Lett.\/} {\bf 75}~(4), 755--758.

\bibitem[Ajdari (2000)]{ajdari00}
{\sc Ajdari, A.} 2000 Pumping liquids using asymmetric electrode arrays. {\em
  Phys. Rev. E\/} {\bf 61}~(1), R45--R48.

\bibitem[Ajdari (2001)]{ajdari01}
{\sc Ajdari, A.} 2001 Transverse electrokinetic and microfluidic effects in
  micropatterned channels: Lubrication analysis for slab geometries. {\em Phys.
  Rev. E\/} {\bf 6501}~(1), art. no. 016301.

\bibitem[Anderson (1985)]{anderson85b}
{\sc Anderson, J.~L.} 1985 Effect of nonuniform zeta potential on particle
  movement in electric-fields. {\em J. Colloid Interface Sci.\/}
  {\bf 105}~(1), 45--54.

\bibitem[Anderson \& Idol (1985)]{anderson85a}
{\sc Anderson, J.~L. \& Idol, W.~K.} 1985 Electroosmosis through pores with
  nonuniformly charged walls. {\em Chem. Eng. Comm.\/} {\bf
  38}~(3-6), 93--106.

\bibitem[Aref (1984)]{aref84}
{\sc Aref, H.} 1984 Stirring by chaotic advection. {\em J. Fluid Mech.\/} {\bf
  143}, 1--21.

\bibitem[Barany {\em et~al.\/} (1998)]{barany98}
{\sc Barany, S., Mishchuk, N.~A. \& Prieve, D.~C.} 1998 Superfast
  electrophoresis of conducting dispersed particles. {\em J. Colloid
  Interface Sci.\/} {\bf 207}~(2), 240--250.

\bibitem[Bazant \& Squires (2004)]{bazant04a}
{\sc Bazant, M.~Z. \& Squires, T.~M.} 2004 Induced-charge electrokinetic
  phenomena:  theory and microfluidic applications {\em Phys. Rev. Lett.\/},
  to appear.

\bibitem[Bazant, Thornton \& Ajdari (2004)]{bazant04b}
{\sc Bazant, M.~Z., Thornton, K. \& Ajdari, A.} 2004 Diffuse-charge dynamics in
  electrochemical systems. {\em preprint\/}.

\bibitem[Beebe {\em et~al.\/} (2002)]{beebe02}
{\sc Beebe, D.~J., Mensing, G.~A. \& Walker, G.~M.} 2002 Physics and
  applications of microfluidics in biology. {\em Annu. Rev. Biomed. Eng.\/}
  {\bf 4}, 261--286.

\bibitem[Ben \& Chang (2002)]{ben02}
{\sc Ben, Y. \& Chang, H.~C.} 2002 Nonlinear smoluchowski slip velocity and
  micro-vortex generation. {\em J. Fluid
   Mech.\/} {\bf 461},
  229--238.

\bibitem[Bikerman (1933)]{bikerman33} {\sc Bikerman, J. J.} 1933
Ionentheorie der elektrosmose, der str\"omungsstr\"ome und der
oberfl\"achenleitf\"ahigkeit. {\em Zeitschrift fur Physikalische
  Chemie A} {\bf 163}, 378--394.

\bibitem[Bikerman (1935)]{bikerman35} {\sc Bikerman, J. J.} 1935
Wissenschaftliche und technische sammelreferate. Die
oberfl\"achenleitf\"ahigkeit und ihre bedeutung.  {\em Kolloid Zeitschrift}
{\bf 72}, 100--108.

\bibitem[Bikerman (1940)]{bikerman40} {\sc Bikerman, J. J.} 1940
  Electrokinetic equations and surfce conductance. A survey of the
  diffuse double layer theory of colloidal solutions. {\em
  Trans. of the Faraday Soc.} {\bf 36}, 154--160.

\bibitem[Bonnefont {\em et~al.\/} (2001)]{bonnefont01}
{\sc Bonnefont, A., Argoul, F. \& Bazant, M.~Z.} 2001 Analysis of diffuse-layer
  effects on time-dependent interfacial kinetics. {\em J.
  Electroanal. Chem.\/} {\bf 500}~(1-2), 52--61.

\bibitem[Brenner (1958)]{brenner58}
{\sc Brenner, H.} 1958 Dissipation of energy due to solid particles suspended
  in a viscous liquid. {\em Phys. Fluids\/} {\bf 1}~(4), 338--346.
  
\bibitem[Brown {\em et~al.\/} (2001)]{brown01}
{\sc Brown, A. B.~D., Smith, C.~G. \& Rennie, A.~R.} 2001 Pumping of water with
  AC electric fields applied to asymmetric pairs of microelectrodes. {\em
  Phys. Rev. E\/} {\bf 6302}~(2), art. no.--016305.

\bibitem[Bruin (2000)]{bruin00}
{\sc Bruin, G. J.~M.} 2000 Recent developments in electrokinetically driven
  analysis on microfabricated devices. {\em Electrophoresis\/} {\bf 21}~(18),
  3931--3951.

\bibitem[Buck (1969)]{buck69}
{\sc Buck, R.~P.} 1969 Diffuse layer charge relaxation at ideally polarized
  electrode. {\em J. Electroanal. Chem.\/} {\bf 23}~(2),
  219--240.

\bibitem[Deryagin \& S. S. Dukhin (1969)]{deryagin69}
{\sc Deryagin, B. V. \& Dukhin, S. S.} 1969 Theory of surface
conductance. {\em Colloid J. USSR} {\bf 31}~(3), 277--283.

\bibitem[Dolnik \& Hutterer (2001)]{dolnik01}
{\sc Dolnik, V. \& Hutterer, K.~M.} 2001 Capillary electrophoresis of proteins
  1999-2001. {\em Electrophoresis\/} {\bf 22}~(19), 4163--4178.
  
\bibitem[A. S. Dukhin \& Murtsovkin (1986)]{dukhin86a}
{\sc Dukhin, A.~S. \& Murtsovkin, V.~A.} 1986 Pair interaction of particles in
  electric-field.  2. Influence of polarization of double-layer of dielectric
  particles on their hydrodynamic interaction in a stationary electric-field.
  {\em Colloid J. USSR\/} {\bf 48}~(2), 203--209.

\bibitem[A. S. Dukhin (1986)]{dukhin86b}
{\sc Dukhin, A.~S.} 1986 Pair interaction of disperse particles in
  electric-field. 3.  Hydrodynamic interaction of ideally polarizable metal
  particles and dead biological cells. {\em Colloid J. USSR\/} {\bf
  48}~(3), 376--381.

\bibitem[S. S. Dukhin (1965)]{dukhin65}
{\sc Dukhin, S.~S.} 1965 Diffusion-electrical theory of electrophoresis. {\em
  XXth Int. Cong. Pure App. Chem., Moscow\/} {\bf
  A72}, 68.
  
\bibitem[S. S. Dukhin \& Semenikhin (1970)]{dukhin70}
{\sc Dukhin, S.~S. \& Semenikhin, V.~N.} 1970 Theory of double layer
  polarization and its influence on the electrokinetic and electro-optical
  phenomena and the dielectric permeability of disperse systems. 
  {\em Colloid J. USSR\/} {\bf 32}~(3), 298--305.

\bibitem[S. S. Dukhin \& Shilov (1974)]{dukhin74}
{\sc Dukhin, S.~S. \& Shilov, V.~N.} 1974 {\em Dielectric phenomena and the
  double layer in disperse systems and polyelectrolytes\/}. New York: Wiley.

\bibitem[S. S. Dukhin \& Shilov (1980)]{dukhin80}
{\sc Dukhin, S.~S. \& Shilov, V.~N.} 1980 Kinetic aspects of electrochemistry
  of disperse systems.  2. Induced dipole-moment and the nonequilibrium
  double-layer of a colloid particle. {\em Adv. Colloid Interface
  Sci.\/} {\bf 13}~(1-2), 153--195.

\bibitem[S. S. Dukhin (1991)]{dukhin91}
{\sc Dukhin, S.~S.} 1991 Electrokinetic phenomena of the 2nd kind and their
  applications. {\em Adv. Colloid Interface Sci.\/} {\bf 35},
  173--196.

\bibitem[S. S. Dukhin (1993)]{dukhin93}
{\sc Dukhin, S.~S.} 1993 Nonequilibrium electric surface phenomena. {\em
  Adv. Colloid Interface Sci.\/} {\bf 44}, 1--134.

\bibitem[Ferry (1948)]{ferry48}
{\sc Ferry, J.~D.} 1948 Frequency dependence of the capacity of a diffuse
  double layer. {\em J. Chem. Phys.\/} {\bf 16}~(7), 737--739.

\bibitem[Figeys \& Pinto (2001)]{figeys01}
{\sc Figeys, D. \& Pinto, D.} 2001 Proteomics on a chip: Promising
  developments. {\em Electrophoresis\/} {\bf 22}, 208--216.

\bibitem[Gajar \& Geis (1992)]{gajar92}
{\sc Gajar, S.~A. \& Geis, M.~W.} 1992 An ionic liquid-channel field-effect
  transistor. {\em J. Electrochem. Soc.\/} {\bf 139}~(10),
  2833--2840.

\bibitem[Gamayunov, Murtsovkin \& A. S. Dukhin (1986)]{gamayunov86}
{\sc Gamayunov, N.~I., Murtsovkin, V.~A. \& Dukhin, A.~S.} 1986 Pair
  interaction of particles in electric-field.  1. Features of hydrodynamic
  interaction of polarized particles. {\em Colloid J. USSR\/} {\bf
  48}~(2), 197--203.

\bibitem[Gamayunov {\em et~al.\/} (1992)]{gamayunov92}
{\sc Gamayunov, N.~I., Mantrov, G.~I. \& Murtsovkin, V.~A.} 1992 Study of flows
  induced in the vicinity of conducting particles by an external
  electric-field. {\em Colloid J. USSR\/} {\bf 54}~(1), 20--23.

\bibitem[Ghosal (2003)]{ghosal03}
{\sc Ghosal, S.} 2003 The effect of wall interactions in capillary-zone
  electrophoresis. {\em J. Fluid Mech.\/} {\bf 491}, 285--300.

\bibitem[Ghowsi \& Gale (1991)]{ghowsi91b}
{\sc Ghowsi, K. \& Gale, R.~J.} 1991 Field-effect electroosmosis. {\em J.
Chromatogr.\/} {\bf 559}~(1-2), 95--101.

\bibitem[Gitlin {\em et~al.\/} (2003)]{gitlin03}
{\sc Gitlin, I., Stroock, A.~D., Whitesides, G.~M. \& Ajdari, A.} 2003 
Pumping based on transverse electrokinetic effects. {\em App. Phys. Lett.\/} {\bf 83}~(7), 1486--1488.

\bibitem[Gonzalez {\em et~al.\/} (2000)]{gonzalez00}
{\sc Gonzalez, A., Ramos, A., Green, N.~G., Castellanos, A. \& Morgan, H.} 2000
  Fluid flow induced by nonuniform ac electric fields in electrolytes on
  microelectrodes. II. A linear double-layer analysis. {\em Phys. Rev.
  E\/} {\bf 61}~(4), 4019--4028.

\bibitem[Happel \& Brenner (1983)]{happel83}
{\sc Happel, J. \& Brenner, H.} 1983 {\em Low Reynolds Number Hydrodynamics\/}.
  The Hague: Martinus Nijhoff Publishers.

\bibitem[Hayes \& Ewing (1992)]{hayes92}
{\sc Hayes, M.~A. \& Ewing, A.~G.} 1992 Electroosmotic flow-control and
  monitoring with an applied radial voltage for capillary zone electrophoresis.
  {\em Anal. Chem.\/} {\bf 64}~(5), 512--516.

\bibitem[Hunter (2000)]{hunter00}
{\sc Hunter, R.~J.} 2000 {\em Foundations of Colloid Science\/}, 2nd edn.
  Oxford: Oxford University Press.

\bibitem[Jeffrey \& Onishi (1981)]{jeffrey81}
{\sc Jeffrey, D.~J. \& Onishi, Y.} 1981 The slow motion of a cylinder next to a
  plane wall. {\em Q. J. Mech. Appl. Math.\/} {\bf 34}, 129--137.

\bibitem[Kornyshev \& Vorotyntsev (1981)]{kornyshev81}
{\sc Kornyshev, A.~A. \& Vorotyntsev, M.~A.} 1981 Conductivity and space-charge
  phenomena in solid electrolytes with one mobile charge carrier species, a
  review with original material. {\em Electrochimica Acta\/} {\bf 26}~(3),
  303--323.

\bibitem[Landers (2003)]{landers03}
{\sc Landers, J.~P.} 2003 Molecular diagnostics on electrophoretic microchips.
  {\em Anal. Chem.\/} {\bf 75}~(12), 2919--2927.

\bibitem[Lee {\em et~al.\/} (1990)]{lee90}
{\sc Lee, C.~S., Blanchard, W.~C. \& Wu, C.~T.} 1990 Direct control of the
  electroosmosis in capillary zone electrophoresis by using an external
  electric-field. {\em Anal. Chem.\/} {\bf 62}~(14), 1550--1552.

\bibitem[Levich (1962)]{levich62}
{\sc Levich, V.~G.} 1962 {\em Physicochemical Hydrodynamics\/}. Englewood
  Cliffs, N.J.: Prentice-Hall, Inc.

\bibitem[Liron \& Blake (1981)]{liron81}
{\sc Liron, N. \& Blake, J.~R.} 1981 Existence of viscous eddies near
  boundaries. {\em J. Fluid Mech.\/} {\bf 107}~(JUN), 109--129.

\bibitem[Liu {\em et~al.\/} (2000)]{liu00}
{\sc Liu, R.~H., Stremler, M.~A., Sharp, K.~V., Olsen, M.~G., Santiago, J.~G.,
  Adrian, R.~J., Aref, H. \& Beebe, D.~J.} 2000 Passive mixing in a
  three-dimensional serpentine microchannel. {\em J. Microelectromech. Syst.\/}
  {\bf 9}~(2), 190--197.

\bibitem[Long \& Ajdari (1998)]{long98}
{\sc Long, D. \& Ajdari, A.} 1998 Symmetry properties of the electrophoretic
  motion of patterned colloidal particles. {\em Phys. Rev. Lett.\/} {\bf
  81}~(7), 1529--1532.

\bibitem[Long {\em et~al.\/}(1999)]{long99}
{\sc Long, D., Stone, H.~A. \& Ajdari, A.} 1999 Electroosmotic flows created by
  surface defects in capillary electrophoresis. {\em J. Colloid Interface
  Sci.\/} {\bf 212}~(2), 338--349.

\bibitem[Lyklema (1991)]{lyklema91}
{\sc Lyklema, J.} 1991 {\em Fundamentals of interface and colloid science\/},
  vol.~2. London: Academic Press.

\bibitem[Macdonald (1954)]{macdonald54}
{\sc Macdonald, J.~R.} 1954 Theory of the differential capacitance of
the double layer in unadsorbed electrolytes. {\em J. Chem.
  Phys.}, {\bf 22}, 1857--1866.

\bibitem[Macdonald (1970)]{macdonald70}
{\sc Macdonald, J.~R.} 1970 Double layer capacitance and relaxation in
  electrolytes and solids. {\em Trans. Faraday Soc.\/} {\bf
  66}~(568), 943--958.

\bibitem[Mpholo {\em et~al.\/} (2003)]{mpholo03}
{\sc Mpholo, M., Smith, C.~G. \& Brown, A. B.~D.} 2003 Low voltage plug flow
  pumping using anisotropic electrode arrays. {\em Sens. Act.
  B\/} {\bf 92}~(3), 262--268.

\bibitem[Murtsovkin (1991)]{murtsovkin91b}
{\sc Murtsovkin, V.~A.} 1991 Criterion for ideal polarizability of conducting
  particles. {\em Colloid J. Russ. Acad. Sci.\/} {\bf 53}~(6), 947--948.

\bibitem[Murtsovkin (1996)]{murtsovkin96}
{\sc Murtsovkin, V.~A.} 1996 Nonlinear flows near polarized disperse particles.
  {\em Colloid J.\/} {\bf 58}~(3), 341--349.

\bibitem[Nadal {\em et~al.\/} (2002{\natexlab{{\em a\/}}})]{nadal02b}
{\sc Nadal, F., Argoul, F., Hanusse, P., Pouligny, B. \& Ajdari, A.}
  2002{\natexlab{{\em a\/}}} Electrically induced interactions between
  colloidal particles in the vicinity of a conducting plane. {\em Phys.
  Rev. E\/} {\bf 65}~(6), art. no.--061409.

\bibitem[Nadal {\em et~al.\/} (2002{\natexlab{{\em b\/}}})]{nadal02a}
{\sc Nadal, F., Argoul, F., Kestener, P., Pouligny, B., Ybert, C. \& Ajdari,
  A.} 2002{\natexlab{{\em b\/}}} Electrically induced flows in the vicinity of
  a dielectric stripe on a conducting plane. {\em Euro. Phys. J.
  E\/} {\bf 9}~(4), 387--399.

\bibitem[O'Brien \& White (1978)]{obrien78}
{\sc O'Brien, R.~W. \& White, L.~R.} 1978 Electrophoretic mobility of a
  spherical colloidal particle. {\em J. Chem. Soc. Faraday II\/} {\bf 74},
  1607–-1626.

\bibitem[O'Brien \& Hunter (1981)]{obrien81}
{\sc O'Brien, R.~W. \& Hunter, R.~J.} 1981 The electrophoretic mobility of
  large colloidal particles. {\em Can. J. Chem.-Rev. Can. Chim.\/} {\bf
  59}~(13), 1878--1887.
  
\bibitem[O'Brien (1983)]{obrien83}
{\sc O'Brien, R.~W.} 1983 The solution of the electrokinetic equations for
  colloidal particles with thin double-layers. {\em J. Colloid Interface
  Sci.\/} {\bf 92}~(1), 204--216.

\bibitem[Overbeek (1943)]{overbeek43}
{\sc Overbeek, J. T.~B.} 1943 Theorie der elektrophorese. der
  relaxationseffekt. {\em Kolloid Beihefte\/} {\bf 54}, 287--364.

\bibitem[Ramos {\em et~al.\/} (1998)]{ramos98}
{\sc Ramos, A., Morgan, H., Green, N.~G. \& Castellanos, A.} 1998 AC
  electrokinetics: a review of forces in microelectrode structures. {\em J.
  Phys. D-Appl. Phys.\/} {\bf 31}~(18), 2338--2353.

\bibitem[Ramos {\em et~al.\/} (1999)]{ramos99}
{\sc Ramos, A., Morgan, H., Green, N.~G. \& Castellanos, A.} 1999 AC
  electric-field-induced fluid flow in microelectrodes. {\em J. Colloid
  Interface Sci.\/} {\bf 217}~(2), 420--422.

\bibitem[Ramos {\em et~al.\/} (2001)]{ramos01}
{\sc Ramos, A., Gonzalez, A., Green, N.~G., Morgan, H. \& Castellanos, A.} 2001
  Comment on ``theoretical model of electrode polarization and AC electroosmotic
  fluid flow in planar electrode arrays". {\em J. Colloid Interface
  Sci.\/} {\bf 243}~(1), 265--266.

\bibitem[Ramos {\em et~al.\/} (2003)]{ramos03}
{\sc Ramos, A., Gonzalez, A., Castellanos, A., Green, N.~G. \& Morgan, H.} 2003
  Pumping of liquids with AC voltages applied to asymmetric pairs of
  microelectrodes. {\em Phys. Rev. E\/} {\bf 67}~(5), art. no. 056302.

\bibitem[Reyes {\em et~al.\/} (2002)]{reyes02a}
{\sc Reyes, D.~R., Iossifidis, D., Auroux, P.~A. \& Manz, A.} 2002 Micro total
  analysis systems. 1.  Introduction, theory, and technology. {\em Anal.
  Chem.\/} {\bf 74}~(12), 2623--2636.

\bibitem[Ristenpart {\em et~al.\/} (2003)]{ristenpart03}
{\sc Ristenpart, W.~D., Aksay, I.~A. \& Saville, D.~A.} 2003 Electrically
  guided assembly of planar superlattices in binary colloidal suspensions. {\em
  Phys. Rev. Lett.\/} {\bf 90}~(12), art. no.--128303.

\bibitem[Russel {\em et~al.\/} (1989)]{russel89}
{\sc Russel, W.~B., Saville, D.~A. \& Schowalter, W.~R.} 1989 {\em Colloidal
  Dispersions\/}. Cambridge: Cambridge University Press.

\bibitem[Schasfoort {\em et~al.\/} (1999)]{schasfoort99}
{\sc Schasfoort, R. B.~M., Schlautmann, S., Hendrikse, L. \& van~den Berg, A.}
  1999 Field-effect flow control for microfabricated fluidic networks. {\em
  Science\/} {\bf 286}~(5441), 942--945.

\bibitem[Scott {\em et~al.\/} (2001)]{scott01}
{\sc Scott, M., Kaler, K. \& Paul, R.} 2001 Theoretical model of electrode
  polarization and ac electroosmotic fluid flow in planar electrode arrays.
  {\em J. Colloid Interface Sci.\/} {\bf 238}~(2), 449--451.

\bibitem[Shilov \& S. S. Dukhin (1970)]{shilov70}
{\sc Shilov, V.~N. \& Dukhin, S.~S.} 1970 Theory of polarization of diffuse
  part of a thin double layer at a spherical particle in an alternating
  electric field. {\em Colloid J. USSR\/} {\bf 32}~(1), 90--95.

\bibitem[Simonov \& S. S. Dukhin (1973)]{simonov73a}
{\sc Simonov, I.~N. \& Dukhin, S.~S.} 1973 Theory of electrophoresis of solid
  conducting particles in case of ideal polarization of a thin diffuse
  double-layer. {\em Colloid J.\/} {\bf 35}~(1), 173--176.

\bibitem[Simonov \& Shilov (1973)]{simonov73b}
{\sc Simonov, I.~N. \& Shilov, V.~N.} 1973 Theory of the polarization of the
  diffuse part of a thin double layer at conducting, spherical particles in an
  alternating electric field. {\em Colloid J.\/} {\bf 35}~(2), 350--353.

\bibitem[Simonov \& Shilov (1977)]{simonov77a}
{\sc Simonov, I.~N. \& Shilov, V.~N.} 1977 Theory of low-frequency
  dielectric-dispersion of a suspension of ideally polarizable
  spherical-particles. {\em Colloid J. USSR\/} {\bf 39}~(5),
  775--780.

\bibitem[Solomentsev {\em et~al.\/} (1997)]{solomentsev97}
{\sc Solomentsev, Y., Bohmer, M. \& Anderson, J.~L.} 1997 Particle clustering
  and pattern formation during electrophoretic deposition: A hydrodynamic
  model. {\em Langmuir\/} {\bf 13}~(23), 6058--6068.
  
\bibitem[Stone \& Samuel (1996)]{stone96}
{\sc Stone, H.~A. \& Samuel, A. D.~T.} 1996 Propulsion of microorganisms by
  surface distortions. {\em Phys. Rev. Lett.\/} {\bf 77}~(19),
  4102--4104.

\bibitem[Stone \& Kim (2001)]{stone01}
{\sc Stone, H.~A. \& Kim, S.} 2001 Microfluidics: Basic issues, applications,
  and challenges. {\em AICHE J.\/} {\bf 47}, 1250--1254.

\bibitem[Stroock {\em et~al.\/} (2000)]{stroock00}
{\sc Stroock, A.~D., Weck, M., Chiu, D.~T., Huck, W. T.~S., Kenis, P. J.~A.,
  Ismagilov, R.~F. \& Whitesides, G.~M.} 2000 Patterning electro-osmotic flow
  with patterned surface charge. {\em Phys. Rev. Lett.\/} {\bf 84}~(15),
  3314--3317.

\bibitem[Stroock {\em et~al.\/} (2002)]{stroock02a}
{\sc Stroock, A.~D., Dertinger, S. K.~W., Ajdari, A., Mezic, I., Stone, H.~A.
  \& Whitesides, G.~M.} 2002 Chaotic mixer for microchannels. {\em Science\/}
  {\bf 295}~(5555), 647--651.

\bibitem[Studer {\em et~al.\/} (2002)]{studer02}
{\sc Studer, V., Pepin, A., Chen, Y. \& Ajdari, A.} 2002 Fabrication of
  microfluidic devices for AC electrokinetic fluid pumping. {\em
  Microelec. Eng.\/} {\bf 61-2}, 915--920.

\bibitem[Takhistov {\em et~al.\/} (2003)]{takhistov03}
{\sc Takhistov, P., Duginova, K. \& Chang, H.~C.} 2003 Electrokinetic mixing
  vortices due to electrolyte depletion at microchannel junctions. {\em J.
  Colloid Interface Sci.\/} {\bf 263}~(1), 133--143.

\bibitem[Taylor (1966)]{taylor66}
{\sc Taylor, G.~I.} 1966 Studies in electrohydrodynamics.  I. Circulation
  produced in a drop by an electric field. {\em Proc. Roy.
  Soc. London A\/} {\bf 291}~(1425), 159--166.

\bibitem[Thamida \& Chang (2002)]{thamida02}
{\sc Thamida, S.~K. \& Chang, H.~C.} 2002 Nonlinear electrokinetic ejection and
  entrainment due to polarization at nearly insulated wedges. {\em Phys.
  Fluids\/} {\bf 14}~(12), 4315--4328.

\bibitem[Trau {\em et~al.\/} (1997)]{trau97}
{\sc Trau, M., Saville, D.~A. \& Aksay, I.~A.} 1997 Assembly of colloidal
  crystals at electrode interfaces. {\em Langmuir\/} {\bf 13}~(24), 6375--6381.

\bibitem[Whitesides {\em et~al.\/} (2001)]{whitesides01a}
{\sc Whitesides, G.~M., Ostuni, E., Takayama, S., Jiang, X. \& Ingber, D.~E.}
  2001 Soft lithography in biology and biochemistry. {\em Annu. Rev. Biomed.
  Eng.\/} {\bf 3}, 335--373.

\bibitem[Whitesides \& Stroock (2001)]{whitesides01b}
{\sc Whitesides, G.~M. \& Stroock, A.~D.} 2001 Flexible methods for
  microfluidics. {\em Phys. Today\/} {\bf 54}~(6), 42--48.

\bibitem[Yeh {\em et~al.\/} (1997)]{yeh97}
{\sc Yeh, S.~R., Seul, M. \& Shraiman, B.~I.} 1997 Assembly of ordered
  colloidal aggregates by electric-field- induced fluid flow. {\em Nature\/}
  {\bf 386}~(6620), 57--59.

\end{thebibliography}

\end{document}